\documentclass[reprint,twocolumn, nofootinbib]{revtex4-2}		

\usepackage{amssymb,amsmath,verbatim,mathtools,needspace,enumitem,etoolbox,graphicx,microtype,afterpage,bm}

\usepackage[dvipsnames, usenames]{xcolor}
\definecolor{linkcolor}{rgb}{0.1,0.2,0.6}
\definecolor{GREEN}{rgb}{0.05,0.5,0.05}
\usepackage[unicode, colorlinks=true, linkcolor=linkcolor, citecolor=linkcolor, filecolor=linkcolor,urlcolor=linkcolor, pdfusetitle]{hyperref}

\usepackage[T1]{fontenc}
\usepackage[utf8]{inputenc}
\usepackage{tabularx, booktabs}
\usepackage[italicdiff]{physics}
\usepackage{float}

\graphicspath{{./Figures/}}

\newcommand{\cuhk}{\affiliation{Department of Physics, The Chinese University of Hong Kong, Hong Kong SAR, China}}

\begin{document}

\title{I-Love-Q Relations of Fermion-Boson Stars}

\author{Kelvin~Ka-Ho~Lam} 
\email[Present address: Illinois Center for Advanced Studies of the Universe \& Department of Physics, University of Illinois Urbana-Champaign, Urbana, Illinois 61801, USA. \\ Email: ]{khlam4@illinois.edu}
\cuhk

\author{Lap-Ming~Lin}
\cuhk
\date{\today}

\begin{abstract}
	\begin{center}
		ABSTRACT
	\end{center}
We investigate the properties of fermion-boson stars (FBSs), which can be viewed as neutron stars with a bosonic dark matter (DM) admixture. A challenge in studying the impact of DM on neutron stars is the absence of a universally accepted nuclear-matter equation of state (EOS), making it difficult to distinguish between the effects of DM and various EOS models. To address this issue, we extend the study of the I-Love-Q universal relations of neutron stars 
to FBSs with a nonrotating bosonic component by solving the Einstein-Klein-Gordon system.
We study how DM parameters, such as the boson particle mass and self-interaction strength, would affect the structure of FBSs and explore the parameter space that leads to deviations from the I-Love-Q relations. 
The properties of FBSs and the level of deviations in general depend sensitively on the DM parameters. For boson particle mass within the range of $\mathcal{O}(10^{-10} \ \mathrm{eV})$, where the Compton wavelength is 
comparable to the Schwarzschild radius of a $1 M_\odot$ star, the deviation is up to about 5\% level if the star contains a few percent of DM admixture. 
The deviation increases significantly with a higher amount of DM.
We also find that the universal relations are still valid to within a 1\% deviation level for boson particle mass $m_b \geq 26.8\times10^{-10} \ \mathrm{eV}$. This effectively sets an upper bound on the boson particle mass, beyond which it becomes not feasible to probe the properties of FBSs by investigating the I-Love-Q relation violations.
\end{abstract}

\maketitle

\section{Introduction}\label{sec:introduction}

Neutron stars (NSs) are one of the densest objects in the universe. To compute their profiles and properties, we need information on their internal structure through an equation of state (EOS), which is still not well understood within the supranuclear density range relevant to NS cores. So far, we still lack a universally accepted EOS for NSs. Instead, there are numerous EOS models predicting different properties of NSs. The observed properties of NSs can thus be used to put constraints on the EOS models~\cite{chatziioannou2015probing,burgio2021neutron}. 	 
NSs not only serve as astrophysical laboratories for studying the supranuclear EOS but also have
the potential to infer the properties of dark matter (DM). The strong gravity and extreme densities make them ideal candidates for exploring the nature of DM particles.

In recent decades, several pieces of evidence supporting the existence of DM have been found, including
observations of galaxy rotation curves \cite{persic1996the,demartino2020dark}, cosmic microwave 
background measurements \cite{padmanabhan2005detecting}, and gravitational lensing \cite{wittman2000detection}. 
While the existence of DM is well established, we still do not know much about the properties of DM 
particles, such as their mass, interaction strength, and even spin-statistics. 
Motivated by particle physics and cosmology, we investigate ultralight bosonic DM particles such as QCD axions \cite{duffy2009axions} and axion-like particles \cite{choi2021recent}, which are potential candidates of DM. Axions are hypothetical particles that, if they exist, could solve the strong CP problem in QCD. 
On the other hand, axion-like particles, suggested by string theory \cite{arvanitaki2010string}, can address cosmological problems such as cosmic inflation. These particles rarely interact with normal matter, and hence they
serve as natural candidates for DM.

NSs with DM admixture have gained a lot of attention in recent years as they may potentially give rise to observational signatures of DM (see \cite{bramante2024dark} for a recent review).  
While there is still no evidence for their existence, these stars have been studied extensively, including their properties \cite{leung2011dark,xiang2014effects,ellis2018dark,leung2022tidal,collier2022tidal}, 
stability conditions \cite{leung2012equilibrium,kain2021dark,routaray2023radial}, and dynamics \cite{emma2022numerical,gleason2022dynamical,reuter2023quasiequilibrium}.  
In this paper, we investigate a subclass of DM-admixed NSs known as fermion-boson stars (FBSs), in 
which the DM component consists of bosonic particles.
These systems were first studied more than thirty years ago~\cite{henriques1989combined,henriques1990combined,jetzer1990stability}. 
More recently, the properties of FBSs, such as their mass-radius relations and tidal deformability, have been investigated \cite{leung2022tidal,Rutherford:2023,Giangrandi:2023,diedrichs2023tidal}. 
Their dynamics and relevance to gravitational wave (GW) observations have also been explored \cite{valdezalvarado2013dynamical,bezares2019gravitational,digiovanni2020dynamical,lee2021could, digiovanni2022can,Nyhan:2022}. 
The renewed interest in these objects stems from the potential implications they hold for our understanding of the properties of bosonic DM. 

Investigating the properties of DM through observations of DM-admixed NSs (if they exist) presents a challenge due to the degeneracy between DM effects and EOS models. 
The presence of DM affects the stellar structure, impacting global quantities of NSs such as the 
mass-radius relation. Since so far there is no universally accepted nuclear matter EOS as mentioned above, one would thus encounter the challenge of distinguishing between the effects of DM and different EOS models, even if the global properties of NSs could be measured with high precision. 
On the other hand, one can also study the properties of axion DM around NSs through their 
electromagnetic signals due to axion-photon couplings within NS magnetospheres 
(e.g., \cite{Prabhu:2021,Noordhuis:2023}).

In the last decade, various approximately EOS-insensitive universal relations connecting different NS properties have been discovered (see \cite{yagi2017approximate,Doneva:2018} for reviews), offering a promising method to probe the DM effects despite our ignorance of the nuclear-matter EOS as we propose in this work. These relations are relatively insensitive to the EOS models at different levels of accuracy. They can be applied to infer physical properties of NSs from observations 
(e.g., \cite{Lau:2010,yagi2013iloveq,kumar2019inferring}). The most robust universal relations are generally insensitive to the EOSs to within about 1\% level over a large range of NS mass \cite{Lau:2010,yagi2013iloveq,Chan:2014}. In particular, the so-called I-Love-Q relations \cite{yagi2013iloveq} relating the moment of inertia, the tidal deformability, and the spin-induced quadrupole moment of slowly rotating NSs have attracted much attention due to their relevance for NS astrophysics and GW physics.

In this paper, we investigate the I-Love-Q relations for FBSs and demonstrate how we may explore the properties of bosonic DM particles, such as their mass and self-interaction strength, by examining the 
parameter space where these universal relations are broken.  
The I-Love-Q relations for DM-admixed NSs have been studied in \cite{cronin2023rotating} using a two-fluid formalism, where the nuclear matter and fermionic DM are described by perfect fluids. 
The tidal deformability of nonrotating FBSs has also been studied using a two-fluid formalism 
\cite{leung2022tidal}, which is applicable only in the limit where the self-interaction of the bosonic DM is strong. In this limit, instead of solving the scalar field equation, the bosonic DM can be described by a perfect fluid with an effective EOS. 
More recently, a more general computation based on the solution of the full set of Einstein-Klein-Gordon system without the assumption of strong self-interaction has been developed \cite{diedrichs2023tidal}. 
In this work, we follow the formulation of \cite{diedrichs2023tidal} to compute the tidal
deformability of nonrotating FBSs.
For the computation of the moment of inertia and spin-induced quadrupole moment, we extend the 
foundational work of Hartle \cite{hartle1967slowly} for a slowly rotating NS by introducing a non-rotating bosonic DM component into the stellar structure. 
Our slowly rotating FBSs thus consist of a rotating fermionic NM component admixed with a nonrotating bosonic DM. Although the DM component does not rotate, it still contributes to the construction of a slowly rotating FBS by influencing the NM through the metric functions. We refer the reader to 
Sec.~\ref{subsec:I} for more discussion about our assumption of nonrotating bosonic DM.

The rest of the article is structured as follows: In Sec.~\ref{sec:background}, we review the model for non-rotating FBSs and discuss their stability conditions. 
In Sec.~\ref{sec:perturbation}, we present the perturbative calculations of the Einstein-Klein-Gordon
system to obtain the moment of inertia, spin-induced quadrupole moment, and tidal deformability of FBSs.
In Sec.~\ref{sec:result}, we present our numerical results and investigate the effects of varying 
bosonic particle masses and self-interaction strength on the I-Love-Q relations. 
In Sec.~\ref{sec:eos}, we study the effects of nuclear-matter EOS on the deviations observed in the
I-Love-Q relations for FBSs. In this work, we use units where $G=c=\hbar=1$ unless otherwise noted.

\section{Background Solution} \label{sec:background}

\subsection{Formulation}

We describe the bosonic component using a complex scalar field $\phi$ with the following self-interacting potential \cite{henriques1989combined}
\begin{align}
	V(\phi)=\frac{1}{2}m_b^2|\phi|^2+\frac{1}{4}\lambda_{\rm int}|\phi|^4, \label{eq:potential}
\end{align}
where $m_b$ is the mass of the scalar boson and $\lambda_{\rm int}$ is the self-interaction constant. A positive (negative) $\lambda_{\rm int}$ means that the
interaction is repulsive (attractive).
The corresponding Lagrangian density for the scalar field in curved spacetime is described by:
\begin{align}
	\mathcal{L}(\phi,\partial_{\mu}\phi)=-\frac{1}{2}g^{\mu\nu}\partial_{\mu}\phi^*\partial_{\nu}\phi-\frac{1}{2}m_b^2|\phi|^2-\frac{1}{4}\lambda_{\rm int}|\phi|^4, \label{eq:Lagrangian}
\end{align}
where $\phi^*$ is the complex conjugate of $\phi$ and 
$g_{\mu\nu}=\mathrm{diag}\,(-e^{2\nu(r)}, e^{2\lambda(r)}, r^2, r^2\sin^2\theta)$ is the metric for 
the spherically symmetric and static background star. The energy-momentum tensor for the scalar field is then 
\begin{align} 
	\begin{split}
		T_{\mu\nu}^{\mathrm{DM}}&=\frac{1}{2}\left(\partial_{\mu}\phi^*\partial_{\nu}\phi+\partial_{\mu}\phi\,\partial_{\nu}\phi^*\right) \\
		&\hspace*{5mm}-\frac{1}{2}g_{\mu\nu}\left(g^{\alpha\beta}\partial_{\alpha}\phi^*\partial_{\beta}\phi+m_b^2|\phi|^2+\frac{1}{2}\lambda_{\rm int}|\phi|^4\right). \label{eq:T_DM}
	\end{split}
\end{align}
In order to have a localized and non-singular boson distribution, we make the ansatz 
\begin{align}
	\phi(r,t)=\Phi(r)e^{-i\gamma t} ,
\end{align}
	where $\gamma$ is a constant to be determined. 
On the other hand, the fermionic normal matter (NM) inside the star is assumed to be a perfect fluid
and described by the following energy-momentum tensor:
\begin{align}
	T_{\mu\nu}^{\mathrm{NM}}=(\epsilon+p)u_{\mu}u_{\nu}+g_{\mu\nu}p, \label{eq:T_NM}
\end{align}
where $p$ and $\epsilon$ are the pressure and total energy density of the fluid, respectively. 
The 4-velocity of the fluid $u^\mu$ is given by $u^{\mu}=(e^{-\nu},0,0,0)$ in the static background star. 
Assuming gravity is the only interaction between DM and NM particles, the Einstein equations read
\begin{align}
	G_{\mu\nu}=8\pi\left(T_{\mu\nu}^{\mathrm{DM}}+T_{\mu\nu}^{\mathrm{NM}}\right).
\end{align}
The equations of motion for the NM and DM are governed by the energy-momentum conservation laws 
$\nabla^\mu T^{\mathrm{NM}}_{\mu\nu}= \nabla^\mu T^{\mathrm{DM}}_{\mu\nu}=0$.
Following~\cite{henriques1989combined,lee2021could}, we also define a set of dimensionless variables for convenience: 
\begin{gather}
	\begin{split}
		x=m_br, \quad \sigma=\sqrt{4\pi}\phi, \quad \Gamma=\frac{\gamma}{m_b}, \\
		\Lambda=\frac{\lambda_{\rm int}}{4\pi m_b^2}, \quad \bar{\epsilon}=\frac{4\pi\epsilon}{m_b^2}, \quad \bar{p}=\frac{4\pi p}{m_b^2} . \label{eq:scaling}
	\end{split}
\end{gather}
The Einstein equations and conservation laws then give
\begin{widetext}
	\begin{align}\label{eq:FBS}
		\begin{split}
			\dv{\lambda}{x}&=\frac{1}{2x}(1-e^{2\lambda})+\frac{xe^{2\lambda}}{2}\left(\Gamma^2e^{-2\nu}+1\right)\sigma^2+\frac{\Lambda}{4}xe^{2\lambda}\sigma^4+\frac{x}{2}\sigma^{\prime\,2}+xe^{2\lambda}\bar{\epsilon} , \\
			\dv{\nu}{x}&=\frac{1}{2x}(e^{2\lambda}-1)+\frac{xe^{2\lambda}}{2}\left(\Gamma^2e^{-2\nu}-1\right)\sigma^2-\frac{\Lambda}{4}xe^{2\lambda}\sigma^4+\frac{x}{2}\sigma^{\prime\,2}+xe^{2\lambda}\bar{p} , \\
			\dv[2]{\sigma}{x}&=-\left(\frac{2}{x}+\dv{\nu}{x}-\dv{\lambda}{x}\right)\dv{\sigma}{x}-e^{2\lambda}\left[\left(\Gamma^2e^{-2\nu}-1\right)\sigma-\Lambda\sigma^3\right] , \\
			\dv{\bar{p}}{x}&=-(\bar{\epsilon}+\bar{p})\dv{\nu}{x}.
		\end{split}
	\end{align}
\end{widetext}
Given an EOS for the NM, we can solve this set of equations to construct a background star. 
We choose to use the APR EOS \cite{akmal1998equation} to describe the NM for the results presented in 
Sec.~\ref{sec:result}. We shall also use the DD2 EOS~\cite{typel2010composition} to study the effects of different NM EOS models in Sec.~\ref{sec:eos}.

To solve the above differential equations, we require suitable boundary conditions to determine a physical solution. At the center of the star $x=0$, we impose a set of regularity conditions to integrate the 
equations outward from some small $x$:
\begin{align}\label{eq:regularity}
	\begin{split}
		\lambda(x)&=\frac{1}{6}\left[(\Gamma^2e^{-2\nu_c}+1)\sigma_c^2+\frac{\Lambda}{2}\sigma_c^4+2\bar{\epsilon}_c\right]x^2 , \\
		\nu(x)&=\nu_c+\frac{1}{2}\nu''_cx^2 , \\
		\sigma(x)&=\sigma_c-\frac{1}{6}\left[(\Gamma^2e^{-2\nu_c}-1)\sigma_c-\Lambda\sigma_c^3\right]x^2, \\
		p(x)&=\bar{p}_c-\frac{1}{2}(\bar{\epsilon}_c+\bar{p}_c)\nu''_cx^2, 
	\end{split}
\end{align}
where the prime denotes $d/dx$ and 
\begin{align*}
	\nu''_c=\frac{1}{3}\left[(2\Gamma^2e^{-2\nu_c}-1)\sigma_c^2-\frac{\Lambda}{2}\sigma_c^4+\bar{\epsilon}_c+3\bar{p}_c\right].
\end{align*} 
Variables with subscript $c$ represent quantities evaluated at the center of the star. For a physical boson distribution, it must be localized and nodeless, i.e., the ground state solution. Hence, the scalar field
should decrease monotonically from the center and $\sigma(\infty)=0$. Also, $\sigma'(0)=0$ should be taken for a smooth distribution of matter. Therefore, the spacetime is asymptotically flat and 
$\lambda(\infty)=\nu(\infty)=0$. This lefts us with two initial conditions $\rho_c$ and $\sigma_c$ that characterize the FBS.

Obtaining an accurate localized distribution of $\sigma(x)$ is quite challenging as it requires to 
determine a precise value of $\Gamma$. We employ a binary search algorithm to compute $\Gamma$. 
If $\Gamma$ is too large, we would obtain a solution of an excited state, where $\sigma$ has nodes along the radius. If $\Gamma$ is small, $\sigma$ is not localized and will diverge to infinity. Hence, we have a binary condition for the algorithm to search for $\Gamma$. To guarantee convergence at the outer boundary so that quantities calculated at the outer boundary are well behaved, we cut off the bosonic tail whenever $\sigma=10^{-4}\sigma_c$ \cite{diedrichs2023tidal}. We define the bosonic radius $r_{\rm DM}$ as the cutoff radius. Outside the cutoff, we simply set $\sigma=0$.

We can define the individual gravitational masses of fermionic and bosonic matter as follows. To calculate the fermionic part, we utilize the corresponding Komar mass integral \cite{lee2021could}
\begin{align}
	M_{\rm NM}=\int_0^{r_{\rm NM}}(\epsilon+3p)e^{2\nu+2\lambda}4\pi r^2\,dr ,
\end{align}
where $r_{\rm NM}$ is the radius of normal fermionic matter. The bosonic part can then be calculated by subtracting the fermionic part from the total mass $M$, expressed as $M_{\rm DM}=M-M_{\rm NM}$, where $M=r_s(1-e^{-2\lambda(r_s)})/2$ and $r_s=\max\,[r_{\rm NM},r_{\rm DM}]$ denotes the surface radius of the star.

Outside of the star, the exterior vacuum solution should satisfy the boundary conditions $\lambda(\infty)=\nu(\infty)=0$. We can first choose any arbitrary value of $\nu(0)$ and integrate the
system outward. Since $\nu(x)$ only appears as $d\nu/dx$ or $\Gamma^2e^{-2\nu}$, we can always transform $\nu\mapsto\nu+\alpha$ to obtain another solution, as long as we also make the corresponding change
$\Gamma\mapsto\Gamma e^{\alpha}$ \cite{henriques1989combined}. This means we can choose any $\nu(0)$ to 
find $\Gamma$ iteratively. After obtaining the correct $\Gamma$, we shift $\nu$ such that $\nu(\infty)=0$. 

\subsection{Dynamical stability}

\begin{figure}[t!]
	\centering
	\includegraphics[width=\columnwidth]{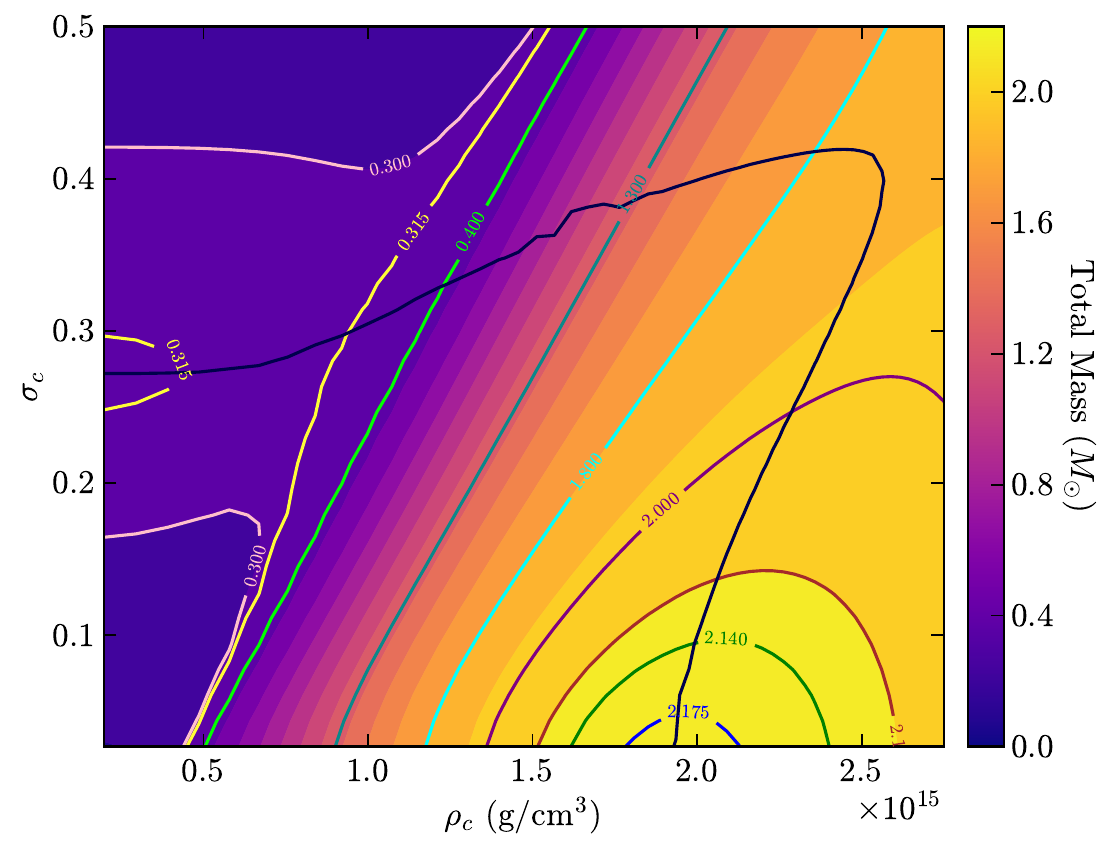}
	\caption{Stability curve (black) of FBSs with $m_b=2.68\times10^{-10}\,\mathrm{eV}$ and $\Lambda=0$. The contours and colorbar represent the total mass of the FBS corresponding to different central densities $\rho_c$ and $\sigma_c$. The stable configurations are located in the lower left region 
		bounded by the stability curve.  }
	\label{fig:stab_curve}
\end{figure}

To investigate the dynamical stability of FBSs, we first introduce two conserved particle numbers corresponding to fermionic and bosonic particles. From Eq.~(\ref{eq:Lagrangian}), we see that the scalar field exhibits global U(1) symmetry that gives rise to a conserved Noether current
\begin{align*}
	j_{\mu}=\frac{i}{2}\left[\phi(\partial_{\mu}\phi^*)-(\partial_{\mu}\phi)\phi^*\right].
\end{align*}
This allows us to define the total number of bosons in the system as
\begin{align}
	N_b&=\int\sqrt{-g}g^{0\mu}j_{\mu}\,d^3r.
\end{align}
Similarly, the particle number for fermions is conserved due to the conservation law
$\nabla_{\mu}(nu^{\mu})=0$, where $n$ is the baryon number density. 
The total number of baryons is defined as
\begin{align}
	N_f&=\int\sqrt{-g}g^{0\mu}nu_{\mu}\,d^3r.
\end{align}

Next, we briefly review the stability criteria for NS and BS. 
For NSs, radial oscillation modes determine the stable and unstable branch along a sequence of nonrotating NSs with the same EOS. When the radial oscillation mode frequency is real (imaginary), the star is stable (unstable). These two branches are separated by a turning point with a zero-frequency mode, meaning that the 
perturbation changes from one equilibrium solution to a nearby one with the same total number of baryon 
and total mass, i.e., $dM_{\rm NM}/d\rho_c = 0$, where $\rho_c$ is the central baryon mass density \cite{shapiro1983black}. 
Similarly, in the case of BS, by the same argument, the turning point at $\sigma_c=\sigma_0$, defined by 
$dM_{\rm DM}(\sigma_0)/d\sigma_c=0$ separates the stable and unstable stars along a sequence of increasing
$\sigma_c$ \cite{lee1989stability}. 
Also, $dN_b/d\sigma_c$ vanishes at $\sigma_0$.

In the case of FBSs, $M$, $N_f$ and $N_b$ now depend on two central densities $\rho_c$ and $\sigma_c$. According to~\cite{henriques1990combined,jetzer1990stability}, the turning points now form a transition line that separates the stable and unstable branch. These turning points satisfy the following conditions:
\begin{align} \label{eq:stability}
	\pdv{N_f}{\xi}=0 \quad\mathrm{and}\quad \pdv{N_b}{\xi}=0, 
\end{align}
where 
\begin{align}
	\pdv{\xi}\propto-\pdv{M}{\sigma_c}\pdv{\rho_c}+\pdv{M}{\rho_c}\pdv{\sigma_c}
\end{align}
denotes the derivative along the direction of constant $M$ \cite{diedrichs2023tidal}. 
In theory, the two conditions give the same transition line that separates the stable and unstable branch. However, in practice, the two lines calculated deviate slightly from each other 
due to numerical resolution. We take the curve corresponding to $\partial N_b/\partial\xi=0$ as the correct solution. 

\begin{figure}[t!]
	\centering
	\includegraphics[width=\columnwidth]{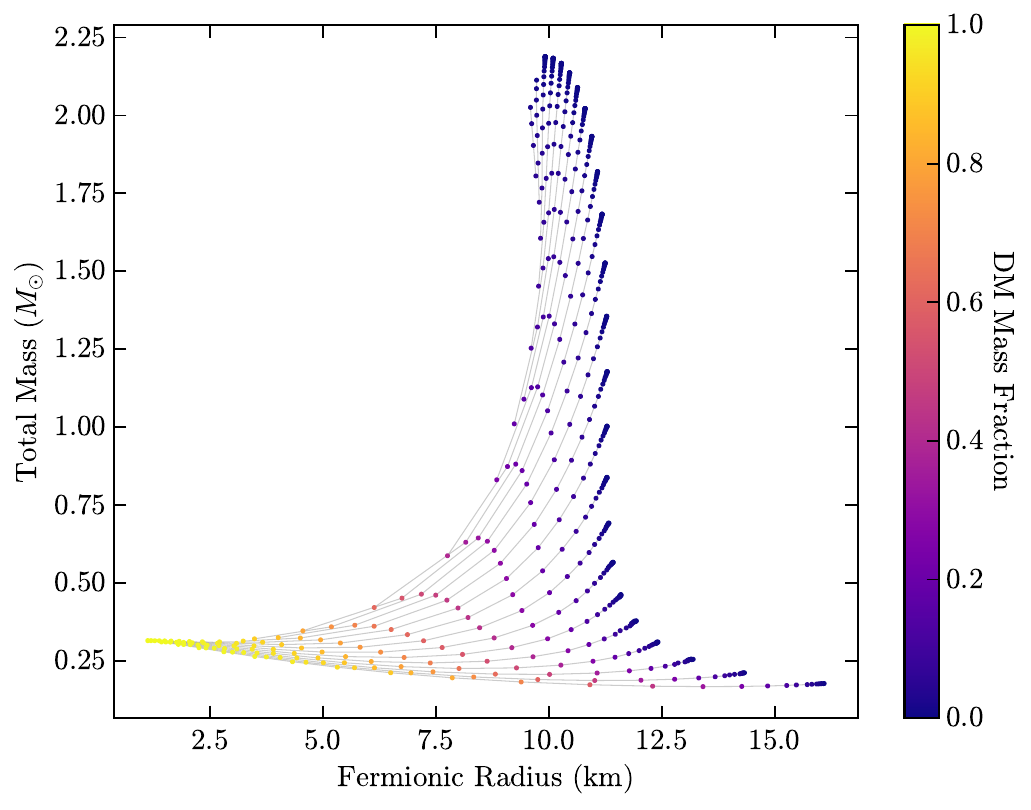}
	\caption{Mass-radius relation of FBSs with $m_b=2.68\times10^{-10}\,\mathrm{eV}$ and $\Lambda=0$. The colorbar represents the DM mass fraction of the FBSs. Configurations with the same $\rho_c$ are connected by thin gray lines.
	}
	\label{fig:mr}
\end{figure}

As an illustration, we determine the region in the parameter space $(\rho_c, \sigma_c)$ within which the 
stars are stable for $m_b=2.68\times 10^{-10}$ eV and $\Lambda=0$ in Fig.~\ref{fig:stab_curve}. 
In the figure, the stable configurations are located in the lower left region bounded by the stability curve (black line), while other regions representing unstable configurations. 
The colorbar represents the total mass of the FBS, and the contours trace configurations with the same masses, 
labeled by the numerical values. 
In the purely fermionic limit ($\sigma_c \rightarrow 0$), the stability curve coincides with the stability turning point (i.e., the maximum mass limit) of NSs for the selected APR EOS. 
Similarly, the purely bosonic limit ($\rho_c \rightarrow 0$) also lies on the turning point of BS. 
The stable configurations for both pure bosonic and fermionic cases must lie in the stable region, indicating that the lower left region is stable.

We also present the mass-radius relation of the stable configurations in Fig.~\ref{fig:mr}. 
In the figure, the colorbar represents the DM mass fraction of the FBSs, and the configurations with the 
same $\rho_c$ are connected by gray lines. The rightmost configuration on each gray line represents a pure
NS, and these data form the mass-radius relation for the APR EOS. 
On the other hand, in the bosonic limit, the FBS becomes a pure BS, as suggested by its maximum stable mass of $0.633/m_b$ \cite{liebling2023dynamical}.

\section{Perturbation Solution} \label{sec:perturbation}

\subsection{Moment of inertia} \label{subsec:I}

To obtain the moment of inertia of a FBS, we extend the perturbative procedure described in \cite{hartle1967slowly,yagi2013iloveq} for slowly rotating NSs to star models with NM admixed with bosonic DM. The NM component is assumed to be slowly rotating. However, we restrict our study to FBSs with non-rotating bosonic components. It should be noted that the assumption of axisymmetric spacetime for a rotating BS imposes a quantized angular momentum $J=aN_b$, where $a$ is an integer \cite{schunck1998rotating}. 
This implies that rotating BSs cannot have arbitrary angular momentum, making perturbation theory
near the nonrotating background fail in the case of BSs, and hence no slowly rotating BSs can be constructed~\cite{kobayashi1994does}. In the limit of no fermionic matter, FBSs will become BSs which cannot rotate perturbatively.
This assumption of non-rotating bosonic matter was also adopted by~\cite{deSousa2001slowly}, which computed the moment of inertia of FBSs. In the following, we focus on FBSs with a non-rotating bosonic component featuring a stationary core or cloud overlapping with the NM. While the bosonic component does not rotate, it still contributes to the construction of a slowly rotating FBS by affecting the metric functions of the spherically symmetric background spacetime.

To construct a slowly rotating star, we consider an axisymmetric and stationary spacetime described by the following metric \cite{yagi2013iloveq,hartle1967slowly}
\begin{widetext}
	\begin{align}\label{eq:perturbation_metric}
		ds^2=-e^{2\nu}(1+2h)dt^2+e^{2\lambda}\left[1+\frac{2m}{r-2\bar{M}}\right]dr^2+r^2(1+2k)[d\theta^2+\sin^2\theta(d\varphi-\omega dt)^2]+\mathcal{O}(\Omega^3), 
	\end{align}
\end{widetext}
where $\bar{M}=r(1-e^{-2\lambda})/2$ is the enclosed mass, $\omega$ is the angular velocity of an inertial frame at $(r,\theta)$ relative to infinity, $\Omega$ is the angular velocity of fermionic
component of the star, assumed to be uniformly rotating, relative to infinity. The metric 
perturbations
\begin{align}
	\begin{split}
		h(r,\theta)&=h_0(r)+h_2(r)P_2(\cos\theta) , \\
		m(r,\theta)&=m_0(r)+m_2(r)P_2(\cos\theta) , \\
		k(r,\theta)&=k_0(r)+k_2(r)P_2(\cos\theta) ,
	\end{split}
\end{align}
are second order in $\Omega$. 
The first-order perturbation is due only to $\omega$ in $g_{t\varphi}$. 
Considering the $t\varphi$ component of the Einstein equations, using dimensionless variables, we have
\begin{align}\label{eq:1st_order}
	\dv[2]{\tilde{\omega}}{x}&=-\left[\frac{4}{x}+\frac{1}{j}\dv{j}{x}\right]\dv{\tilde{\omega}}{x}+4e^{2\lambda}(\bar{\epsilon}+\bar{p})\tilde{\omega}, 
\end{align}
where $\tilde{\omega}=\Omega-\omega$ and $j=e^{-(\nu+\lambda)}$. 
The same equation is also derived in~\cite{deSousa2001slowly}.
It should be noted that for ordinary NSs, the equivalent equation for Eq.~(\ref{eq:1st_order}) can be simplified using the Tolman-Oppenheimer-Volkoff (TOV) equations for the unperturbed background quantities~\cite{yagi2013iloveq,hartle1967slowly}. While in our case for FBSs, the TOV equations are not valid as we also need to solve for the bosonic component. The same happens for Eqs.~(\ref{eq:h2p})~and~(\ref{eq:vp}) below.

Similar to the nonrotating background equations, when $x\to0$, we find the regularity condition 
to be 
\begin{align}
	\tilde{\omega}(x)=\tilde{\omega}_c+\frac{2}{5}(\bar{\epsilon}_c+\bar{p}_c)\tilde{\omega}_cx^2.
\end{align}
The exterior solution is 
\begin{align}\label{eq:exterior_omega}
	\tilde{\omega}=\Omega-\frac{2J}{x^3}, 
\end{align}
where $J$ is the angular momentum of the star. Furthermore, the moment of inertia $I$ is defined
by
\begin{align}
	I=\frac{J}{\Omega}=\frac{x_s^4}{6\Omega}\left(\dv{\tilde{\omega}}{x}\right)_{x=x_s},
\end{align}
where $x_s=m_br_s$ is the dimensionless surface radius. Notice $I$ depends only on the ratio 
$\tilde{\omega}/\Omega$, and Eq.~(\ref{eq:1st_order}) is linear in $\tilde{\omega}$. Thus, $I$ is independent of the angular velocity of the star. In practice, we set $\Omega = 1$ for all computation. We also define the dimensionless moment of inertia $\bar{I}=I/M^3$, which is one of the quantities in the I-Love-Q trio.

\subsection{Spin-induced quadrupole moment} \label{subsec:Q}

The spin-induced quadrupole moment of the star is determined by studying perturbation theory up to 
second order in $\Omega$.  	
The perturbed Einstein equations yield the following equations for calculating the quadrupole moment 
$Q$ \cite{yagi2013iloveq,hartle1967slowly}:
\begin{widetext}
	\begin{align}
		\dv{h_2}{x}&=\left\{-2\nu'+\frac{x}{(x-2\bar{M})\nu'}\left[(\bar{\epsilon}+\bar{p})-\frac{2\bar{M}}{x^3}\right]\right\}h_2-\frac{2v_2}{x\nu'(x-2\bar{M})} \nonumber \\
		&\hspace*{5mm}+\frac{1}{6}\left[\nu'x-\frac{1}{2(x-2\bar{M})\nu'}\right]x^3j^2(\tilde{\omega}')^2+\frac{2}{3}(\bar{\epsilon}+\bar{p})e^{-2\nu}\left[\nu'x+\frac{1}{2(x-2\bar{M})\nu'}\right]x^3\tilde{\omega}^2 , \label{eq:h2p}\\
		\dv{v_2}{x}&=-2\nu'h_2+x^4\left(\frac{1}{x}+\nu'\right)\left[\frac{2}{3}(\bar{\epsilon}+\bar{p})e^{-2\nu}\tilde{\omega}^2+\frac{1}{6}j^2(\tilde{\omega}')^2\right], \label{eq:vp}
	\end{align}
\end{widetext}
where $v_2=h_2+k_2$ and $'=d/dx$. Other metric perturbation terms can be solved as well, but they are not required in the calculation of $Q$, hence we do not present them here. Notice the system is a set of linear differential equations, where the solution is given by a linear combination of the homogeneous and particular solutions. Therefore, the full solution can be written in the following form: 
\begin{align*}
	h_2(x)= {\tilde c} h_{2h}(x)+h_{2p}(x), \\
	v_2(x)={\tilde c} v_{2h}(x)+v_{2p}(x), 
\end{align*}
where ${\tilde c}$ is a constant. The set of particular solutions $(h_{2p},v_{2p})$ satisfies Eqs.~(\ref{eq:h2p})~and~(\ref{eq:vp}), and the set of homogeneous solutions $(h_{2h},v_{2h})$ satisfies 
\begin{align}
	\dv{h_2}{x}&=\left\{-2\nu'x+\frac{x}{(x-2\bar{M})\nu'}\left[(\bar{\epsilon}+\bar{p})-\frac{2\bar{M}}{x^3}\right]\right\}h_2 \nonumber \\
	&\hspace*{5mm}-\frac{2v_2}{x\nu'(x-2\bar{M})} , \label{eq:h2h} \\
	\dv{v_2}{x}&=-2\nu'h_2. \label{eq:vh}
\end{align}	
To integrate outward from the center $x=0$, we use the following regularity conditions:
\begin{gather*}
	h_{2h}(x)=a_hx^2 \quad\mathrm{and}\quad h_{2p}(x)=a_px^2 , \\
	v_{2h}(x)=b_hx^4 \quad\mathrm{and}\quad v_{2p}(x)=b_px^4 , \\
	b_h=-\frac{1}{2}\nu''_ca_h, \quad
	b_p=b_h+\frac{1}{6}(\bar{\epsilon}_c+\bar{p}_c)e^{-2\nu_c}\tilde{\omega}_c^2 , 
\end{gather*}
where $a_h$ and $a_p$ are arbitrary constants. 		
On the other hand, outside of the star, Eqs.~(\ref{eq:h2p}) and (\ref{eq:vp}) reduce to  
\begin{align}
	\dv{h_2}{x}&=-\frac{2v_2}{M}-\frac{2(x-M)}{x(x-2M)}h_2-\frac{3J^2}{M}\frac{x^2-2Mx-2M^2}{x^5(x-2M)} \label{eq:h2_ext} , \\
	\dv{v_2}{x}&=-\frac{2M}{x(x-2M)}h_2+\frac{6J^2}{x^5}\frac{x-M}{x-2M}. \label{eq:v_ext}
\end{align}
The complete (i.e., homogeneous $+$ particular) exterior solution is
\begin{align}\label{eq:2nd_order_sol}
	\begin{split}
		h_2(x)&=AQ_2^2\left(\frac{x}{M}-1\right)+J^2\left(\frac{1}{Mx^3}+\frac{1}{x^4}\right) , \\
		v_2(x)&=\frac{2AM}{\sqrt{x(x-2M)}}Q_2^1\left(\frac{x}{M}-1\right)-\frac{J^2}{x^4}, 
	\end{split}
\end{align}
where $A$ is an integration constant, and $Q_{\nu}^{\mu}(\zeta)$ is the associated Legendre function of the second kind:
\begin{align*}
	Q_2^1(\zeta)&=\sqrt{\zeta^2-1}\left[\frac{3\zeta^2-2}{\zeta^2-1}-\frac{3}{2}\zeta\ln(\frac{\zeta+1}{\zeta-1})\right] , \\
	Q_2^2(\zeta)&=\frac{3(\zeta^2-1)}{2}\ln(\frac{\zeta+1}{\zeta-1})-\frac{3\zeta^3-5\zeta}{\zeta^2-1} .
\end{align*}	
By numerically integrating outward from the center using arbitrary values of $a_h$ and $a_p$, we 
can match the interior solution with Eq.~(\ref{eq:2nd_order_sol}) at the surface of the star $x_s$ to compute the constant $A$ by the following matching formulae:
\begin{gather*}
	Ah_{2h}^{\mathrm{ext}}(x_s)+h_{2p}^{\mathrm{ext}}(x_s)=Bh_{2h}^{\mathrm{int}}(x_s)+h_{2p}^{\mathrm{int}}(x_s) , \\
	Av_{2h}^{\mathrm{ext}}(x_s)+v_{2p}^{\mathrm{ext}}(x_s)=Bv_{2h}^{\mathrm{int}}(x_s)+v_{2p}^{\mathrm{int}}(x_s) . 
\end{gather*}
Solving this system of equations gives
\begin{align*}
    A=\left[\frac{v_{2p}^{\mathrm{ext}}(x_s)-v_{2p}^{\mathrm{int}}(x_s)}{v_{2h}^{\mathrm{int}}(x_s)}-\frac{h_{2p}^{\mathrm{ext}}(x_s)-h_{2p}^{\mathrm{int}}(x_s)}{h_{2h}^{\mathrm{int}}(x_s)}\right] \\
    \hspace*{5mm}\times\left[\frac{h_{2h}^{\mathrm{ext}}(x_s)}{h_{2h}^{\mathrm{int}}(x_s)}-\frac{v_{2h}^{\mathrm{ext}}(x_s)}{v_{2h}^{\mathrm{int}}(x_s)}\right]^{-1}.
\end{align*}
The spin-quadrupole moment $Q$ can then be determined by comparing the coefficient of the
$P_2(\cos\theta)/r^3$ term in the perturbed metric $\Phi=h_0+h_2P_2(\cos\theta)$ with the Newtonian potential in the far field limit
\begin{align*}
	\Phi(r,\theta)=-\frac{M}{r}P_0(\cos\theta)-\frac{Q}{r^3}P_2(\cos\theta)+\cdots.
\end{align*}
As a result, we have
\begin{align}
	Q = -\frac{J^2}{M}-\frac{8}{5}AM^3.
\end{align}
Notice $Q<0$ indicates that the star is an oblate spheroid. We also define the dimensionless spin quadrupole moment $\bar{Q}=-QM/J^2$, which is the second quantity in the I-Love-Q trio.

\subsection{Tidal deformability} \label{subsec:lambda}

The tidal deformability $\lambda_{\rm tidal}$ of a star in a binary system measures its deformation due to the tidal field produced by the companion. The computation of $\lambda_{\rm tidal}$ for 
ordinary NSs is well established (e.g., \cite{hinderer2008tidal,yagi2013iloveq}). The formulation for FBSs has recently been studied in \cite{diedrichs2023tidal}, from which we outline and summarize the 
relevant equations below.

For a nonrotating star in the static-tide limit, the quadrupolar static tidal field is described by 
the following metric perturbations in the Regge-Wheeler gauge:   
\begin{align}\label{eq:h_munu_tidal}
	\begin{split}
		h_{\mu\nu}&=Y_{20}(\theta,\varphi)\,\mathrm{diag}\Big(\!-e^{2\nu(r)}H_0(r),e^{2\lambda(r)}H_2(r), \\
		&\hspace*{36mm} r^2K(r),r^2\sin^2\theta K(r)\Big) ,
	\end{split}
\end{align}
where $Y_{20}(\theta, \varphi)$ is the $(l=2, m=0)$ spherical harmonic function. The perturbed 
Einstein equations give
\begin{align*}
	\delta G_{\mu\nu}=8\pi(\delta T_{\mu\nu}^{\rm NM}+\delta T_{\mu\nu}^{\rm DM}) ,
\end{align*}
where the perturbed energy-momentum tensor for the NM is $\delta T_{\mu\nu}^{\rm NM}=\delta p\cdot\,\mathrm{diag}\,(-1/c_s^2,1,1,1)$ and $c_s^2=d\bar{p}/d\bar{\epsilon}$. For the DM part 
$\delta T_{\mu\nu}^{\rm DM}$, we take the first-order terms from Eq.~(\ref{eq:T_DM}), which are linear in the static perturbation of scalar field 
\begin{align*}
	\delta\phi=\frac{\phi_1(r)}{r}Y_{20}(\theta,\varphi).
\end{align*}
The relevant equations for determining the tidal deformability of a nonrotating FBS are 
\cite{diedrichs2023tidal}
\begin{widetext}
	\begin{gather}
		\begin{split}
			\sigma''_1+(\nu'-\lambda')\sigma'_1&+\bigg[2\sigma'+x\sigma''-(\nu'+\lambda')x\sigma'-\Gamma^2x\sigma e^{2\lambda-2\nu}\bigg]H_0 \\
	&-\left[\frac{6e^{2\lambda}}{x^2}+\frac{\nu'-\lambda'}{x}+4\sigma^{\prime\,2}+e^{2\lambda}\left(1+3\Lambda\sigma^2-\Gamma^2e^{-2\nu}\right)\right]\sigma_1=0 , \label{eq:sig_1}
		\end{split} \\
		\begin{split}
			&H''_0+\left[\nu'-\lambda'+\frac{2}{x}\right]H'_0+\left[-\frac{1+3c_s^2}{c_s^2}\sigma^{\prime\,2}+\Gamma^2e^{2\lambda-2\nu}\frac{c_s^2-1}{c_s^2}\sigma^2-2\nu'(\lambda'+\nu')+2\nu''+\frac{3\lambda'+7\nu'}{x}+\frac{\lambda'+\nu'}{xc_s^2}-\frac{6e^{2\lambda}}{x^2}\right]H_0 \\
			&=\frac{2}{x}\left[-\frac{1+3c_s^2}{c_s^2}\sigma''+\left(3\lambda'+\nu'+\frac{\lambda'-\nu'}{c_s^2}-\frac{2}{x}\frac{1+3c_s^2}{c_s^2}\right)\sigma'+e^{2\lambda}\left(\big(1+\Lambda\sigma^2\big)\frac{c_s^2+1}{c_s^2}+\Gamma^2e^{-2\nu}\frac{c_s^2-1}{c_s^2}\right)\sigma\right]\sigma_1. \label{eq:H_0}
		\end{split}
	\end{gather}
\end{widetext}
Eq.~(\ref{eq:sig_1}) results from the linearized equation of motion for the scalar field, while 
Eq.~(\ref{eq:H_0}) is derived from the linearized Einstein equations. 
Comparing to \cite{diedrichs2023tidal}, the above equations are rewritten using our metric convention, normalization condition of $\phi$ and dimensionless parameters defined in Eq.~(\ref{eq:scaling}). Note that we also define $\sigma_1=\sqrt{4\pi}\phi_1$ and primed variables should be understood as derivatives with respect to $x$. 	
One can also check that Eq.~(\ref{eq:H_0}) reduces to the corresponding equation (see Eq. (15)
of \cite{hinderer2008tidal}) for an ordinary NS when the scalar field vanishes.

To solve the above set of differential equations, we put a series expansion about the center of the star into the equations and get
\begin{align*}
	\sigma_1(x)=\sigma_{1c}x^3 , \\
	H_0(x)=H_{0c}x^2.
\end{align*}
Notice Eqs.~(\ref{eq:sig_1})~and~(\ref{eq:H_0}) are linear in both $\sigma_1$ and $H_0$, so we can freely scale both variables such that we fix $H_{0c}=1$. To find $\sigma_{1c}$, we use a similar procedure as finding $\Gamma$. We can perform a bisection algorithm on $\sigma_{1c}$ according to its divergent behavior at infinity. We then search for the solution with no nodes, i.e., the ground state solution. After convergence, we set the cutoff of $\sigma_1$ to be the same as $\sigma$. That is,   
$\sigma_1(x)=0$ when $x\geq mr_{\rm DM}$.

Once the interior solution of $H_0(x)$ is obtained, the remaining step is the same as that for an 
ordinary NS \cite{hinderer2008tidal}. 
One needs to match the interior solution to the exterior solution, which is determined by solving 
Eq.~(\ref{eq:H_0}) in vacuum:  
\begin{align*}
	H''_0+\frac{2(x-M)}{x-2M}H'_0+\frac{6x^2-12Mx+4M^2}{x^2(x-2M)^2}H_0&=0.
\end{align*}
We refer the reader to \cite{diedrichs2023tidal} for more information, and simply 
present the final formula for us to calculate $\lambda_{\rm tidal}$: 
\begin{align}
	\begin{split}
		\lambda_{\rm tidal}&=\frac{16M^5}{15}(1-2C)^2(2+2C(y-1)-y)\,\times \\
		&\hspace*{5mm}\Big[3(1-2C)^2(2-y+2C(y-1))\ln(1-2C) \\
		&\hspace*{6mm}+2C(6-3y+3C(5y-8)) \\
		&\hspace*{6mm}+4C^3(13-11y+C(3y-2)+2C^2(1+y))\Big]^{-1}, 
	\end{split}
\end{align}
where $C=M/x_s$ and $y=x_sH'_0(x_s)/H_0(x_s)$. In the I-Love-Q relation, we use the dimensionless tidal deformability $\bar{\lambda}_{\rm tidal}=\lambda_{\rm tidal}/M^5$ instead.

\section{Result} \label{sec:result}
	
In our study, we choose the mass of the scalar boson $m_b\sim \mathcal{O}(10^{-10} \ \mathrm{eV})$ as 
the boson's Compton wavelength in this mass range is roughly equal to the Schwarzschild radius of a $1M_{\odot}$
star \cite{diedrichs2023tidal, digiovanni2022can}. 
This choice implies that the mass of the bosonic component of a FBS can be as large as $\mathcal{O}(1 M_{\odot})$. We only study and present the results for stable stars, and do not consider unstable configurations in the following.

\subsection{I-Love-Q relations} \label{subsec:I_Love_Q}

\begin{figure*}[t]
	\centering
	\includegraphics[width=\textwidth]{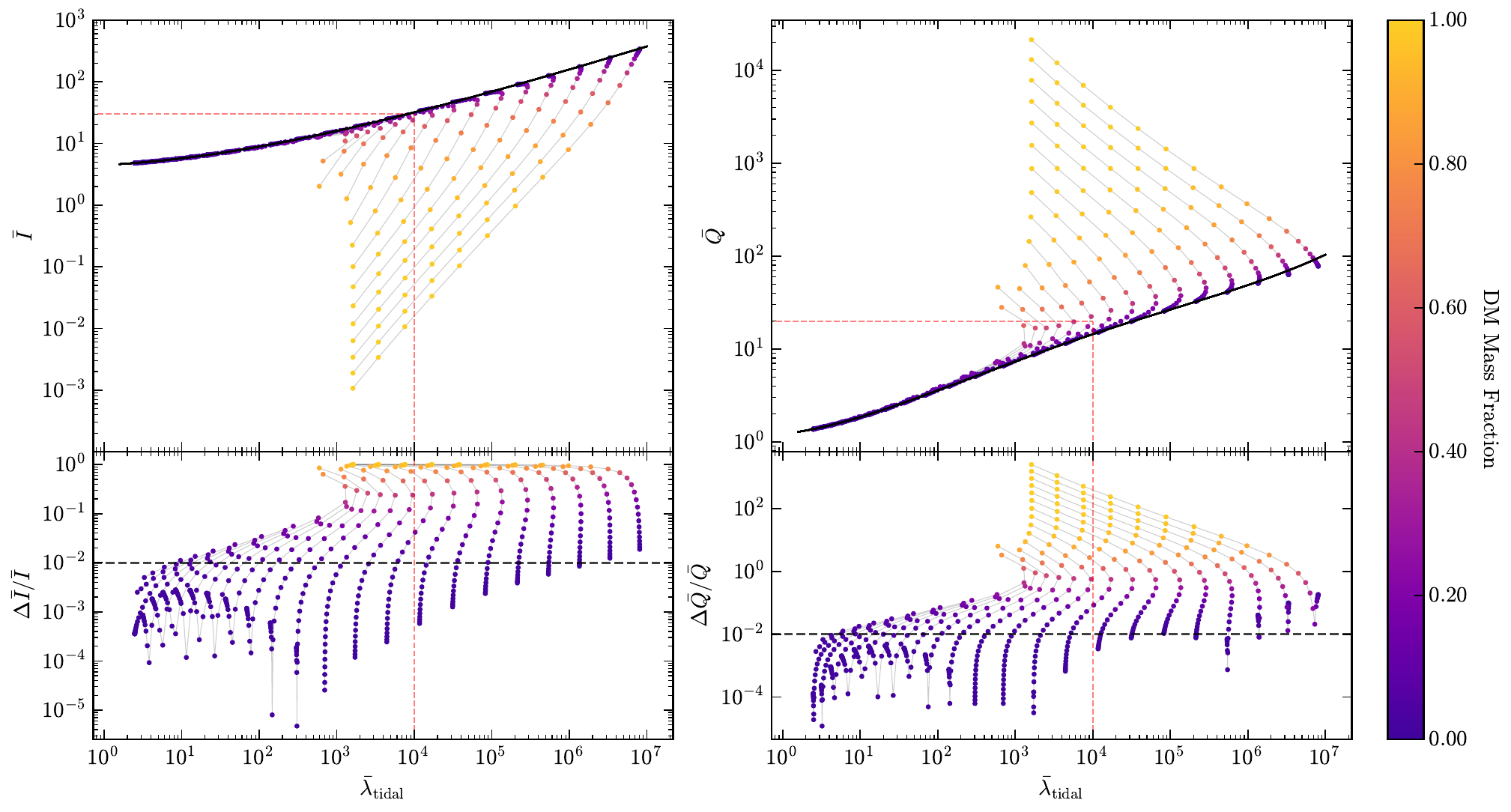}
	\caption{Top panels: I-Love (left) and Q-Love (right) relations for FBSs with $m_b=2.68\times10^{-10}\,\mathrm{eV}$ and $\Lambda=0$. Bottom panels: The fractional errors between the FBS data and the corresponding fitting curves (black solid lines in the top panels)
		for NSs found in Ref.~\cite{yagi2017approximate}. Note that the fitting curves are insensitive to NS 
		EOS models to within $1\%$ in the range $\bar{I}<30$, $\bar{Q}<20$ and $\bar{\lambda}_{\rm tidal}<10^4$, which is bounded by the red dashed lines in the top panels. 
		Different FBSs with the same $\rho_c$ are connected by thin gray lines, while the colorbar represents the DM mass fraction of each FBS.} 
	\label{fig:ILoveQ}
\end{figure*}

The I-Love-Q relations for ordinary NSs are insensitive to many EOS models to within 1\% level \cite{yagi2013iloveq}. We first investigate the effects of the I-Love-Q relations with the introduction of scalar bosonic fields.   
In the top panel of Fig.~\ref{fig:ILoveQ} (left), we plot $\bar I$ against ${\bar \lambda}_{\rm tidal}$ (i.e., the I-Love relation) for sequences of FBSs with fixed boson particle mass $m_b=2.68\times 10^{-10}$ eV and (normalized) coupling constant $\Lambda=0$. 
Each sequence of stars constructed with different DM mass fraction $f=M_{\rm DM}/M$ (represented by 
different colors), but with the same $\rho_c$, is connected by a thin gray line. 
The range of $\rho_c$ covered by the gray lines begins at $\rho_c= 3.16\times 10^{14} \ \mathrm{g/cm}^3$ and ends at the critical value determined by the stability analysis (see Fig.~\ref{fig:stab_curve}).	         
The black solid line is the fitting curve for ordinary NSs \cite{yagi2013iloveq}. 
Similarly, we plot $\bar Q$ against ${\bar \lambda}_{\rm tidal}$ (i.e., the Q-Love relation)
for the same set of FBS data in the top panel of Fig.~\ref{fig:ILoveQ} (right), where the  
corresponding fitting curve for ordinary NSs is still represented by a black solid line. 
The relative errors between FBS data and the fitting curves for NSs are shown in the bottom panels 
of Fig.~\ref{fig:ILoveQ}. 
It should be noted that the original I-Love (Q-Love) fitting curve is determined from NS data in the range 
${\bar I} < 30$ ($\bar{Q} < 20$) and ${\bar \lambda}_{\rm tidal} < 10^4$, which are represented by the red dashed lines
in the figure. In our study of FBSs, we extrapolate the fitting curve to a larger range to cover 
our FBS data.

For FBSs with a small amount of DM ($f\lesssim 5\%$), they can still be fitted by the I-Love-Q relations for NSs very well at a level of about $1\%$, meaning that the universality remains unaffected. 
This is not surprising as FBSs are very similar to pure NSs when fermionic matter dominates. The 
bosonic component has little effect on the I-Love-Q relations. 
For a slightly larger DM mass fraction ($f \sim10\%$), the FBS data start to deviate from the universal relations. 
Specifically, in the I-Love relation, the FBS data lie below the fitting curve, indicating 
that the dimensionless moment of inertia $\bar I$ of FBS is lower than that of a pure NS for the same 
${\bar \lambda}_{\rm tidal}$. This can be understood by the fact that we have introduced a 
non-rotating bosonic component that lowers the angular momentum $J$ of the star, and hence $\bar{I}$.
On the other hand, in the Q-Love relation, the FBS data shift upwards away from the fitting curve. 
While the spin-quadrupole moment $Q$ decreases as we increase the bosonic content, the dimensionless spin-quadrupole moment $\bar{Q}=-QM/J^2$ in fact increases as $J$ also decreases.

When $f\approx0.5$, the constant $\rho_c$ lines (gray lines) exhibit turning points, beyond which the 
data deviate significantly from the I-Love-Q relations for NSs. 
This change occurs when $M_{\rm DM}$ becomes comparable to $M_{\rm NM}$, which is apparent from their purple to reddish color ($f\sim 50\%$).

One might think that the turning points also indicate a change in the structure of the bosonic component, transitioning from a core-like structure (i.e., bosonic matter lies inside the NS) to a cloud-like structure (i.e., NS lies inside bosonic matter), which should involve a kink when the transition occurs. We note that this is not the cause of the turning point. 
Detailed examples and explanation are given in Secs.~\ref{subsec:structure}~and~\ref{subsec:boson_mass}

When DM dominates, FBSs essentially become pure non-rotating BSs represented by the 
yellow data points in Fig.~\ref{fig:ILoveQ}. 
Regardless of the value of $\rho_c$, $\bar{\lambda}_{\rm tidal}$ recovers the result of BS \cite{diedrichs2023tidal,sennett2017distinguishing}, showing that the fermionic matter 
contributes little effect to the calculation of $\bar{\lambda}_{\rm tidal}$. 
However, the tiny amount of fermions contributes to the star's $\bar{I}$ and $\bar{Q}$, making 
them non-zero. 
We see in Fig.~\ref{fig:ILoveQ} that the constant $\rho_c$ lines are nearly parallel to each other instead of overlapping when DM dominates, meaning that those FBSs are not identical. The lines with lower $\rho_c$ values correspond to smaller values of $\bar{I}$ but larger values of $\bar{Q}$. 
This difference in $\bar{I}$ and $\bar{Q}$, although tiny, indicates the difference of fermionic 
content among the stars. 
It is also seen that the relative difference $\Delta {\bar I}/{\bar I}$ approaches unity when
DM dominates, where $\bar I$ in the denominator is the value obtained from the original I-Love relation for NSs. In this limit, the moment of inertia of a FBS approaches zero, and hence $\Delta {\bar I}/{\bar I} \rightarrow 1$, as there is no angular momentum associated to the nonrotating DM component
\footnote{While the moment of inertia
of a nonrotating NS in general relativity can still be defined formally by considering the slow rotation limit perturbatively, this is not the case for pure BS as the angular momentum of a rotating BS is quantized as we mentioned in Sec.~\ref{subsec:I}.}. 
On the other hand, since $\bar Q$ increases with the amount of DM, the
relative difference $\Delta {\bar Q}/{\bar Q}$ can in general be larger than unity.

\subsection{Star structure} \label{subsec:structure}

\begin{figure*}[t]
	\centering
	\includegraphics[width=\textwidth]{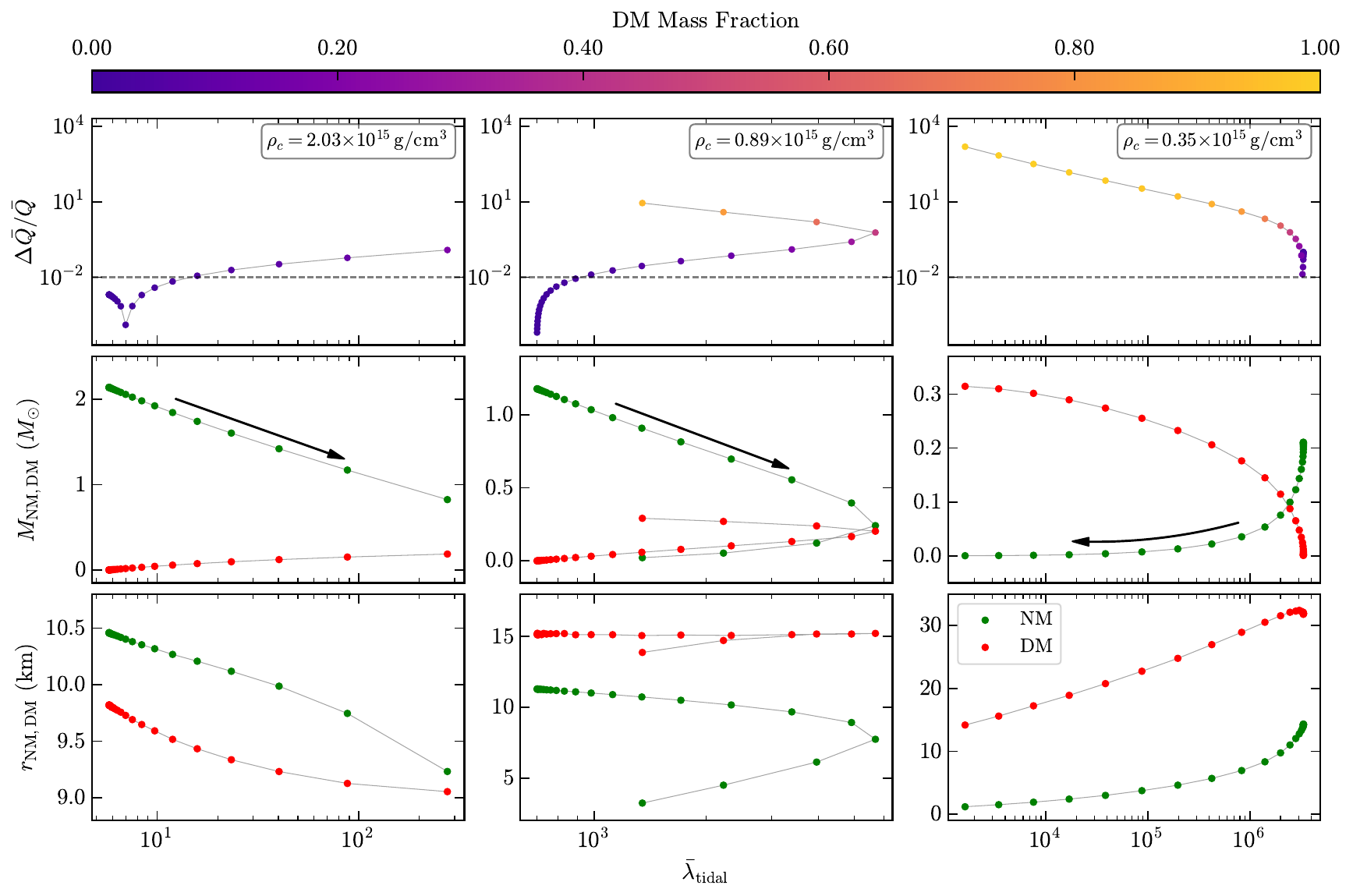}
	\caption{Three sequences of FBSs with $m_b=2.68\times10^{-10}\,\mathrm{eV}$, $\Lambda=0$ 
		and $\rho_c=\{2.03,0.89,0.35\}\times10^{15}\,\mathrm{g/cm}^3$ (from left to right column). 
		Top row: The fractional errors between the FBS data and the Q-Love relation for NSs, with horizontal dashed lines indicating the 1\% EOS-insensitive level as in Fig.~\ref{fig:ILoveQ}.
		The color of each data point represents the DM fraction of the star, as defined by the colorbar.  
		Middle row: Trends of the fermionic NM mass $M_{\rm NM}$ (green data points) and bosonic DM mass $M_{\rm DM}$ (red data points), with arrows indicating the direction of increasing $\sigma_c$. 
		Bottom row: Trends of the fermionic NM and bosonic DM radii. 
		The quantities in all panels are plotted against the dimensionless tidal deformability
		${\bar \lambda}_{\rm tidal}$, but the panels in different columns do not have the same range of
		${\bar \lambda}_{\rm tidal}$.  
	}				
	\label{fig:structure}
\end{figure*}

In this subsection, we focus on the correlation between the structure of FBSs and the breaking of the 
universal relations. In Fig.~\ref{fig:structure}, we consider three sequences of FBSs with constant 
$\rho_c$ to investigate the change of the stellar structure as we increase $\sigma_c$. 
The parameters for the boson particles $m_b=2.68\times 10^{-10}$ eV and $\Lambda=0$ are fixed for all sequences, and this choice is the same as that used for the results presented in Fig.~\ref{fig:ILoveQ}.

In the left column of Fig.~\ref{fig:structure}, we show the results for a sequence of high central 
NM density $\rho_c=2.03\times 10^{15}\ {\rm g/cm}^3$. 
The top panel shows the fractional error $\Delta {\bar Q}/{\bar Q}$ of the Q-Love relation. Similar to
Fig.~\ref{fig:ILoveQ}, the DM fraction along the sequence of FBSs is represented by different colors
(see the colorbar at the top of the figure). We do not show the fractional error in $\bar I$ as the behavior is similar to that of $\bar{Q}$. 
The middle panel shows the trends of the masses of NM $M_{\rm NM}$ (green data points) and DM 
$M_{\rm DM}$ (red data points) as $\sigma_c$ increases along the sequence, with the direction 
indicated by the arrow. The trends of the radii of NM $r_{\rm NM}$ and DM $r_{\rm DM}$ for the 
same sequence are shown in the bottom panel. Note that all the quantities are plotted against 
${\bar \lambda}_{\rm tidal}$. 
Similarly, the results for a sequence of moderate $\rho_c = 0.89\times 10^{15}\ {\rm g/cm}^3$ 
and small $\rho_c=0.35\times 10^{15}\ {\rm g/cm}^3$ are plotted in the middle and right columns, 
respectively.

For any constant $\rho_c$, we find that $M_{\rm DM}$ increases with $\sigma_c$ while $M_{\rm NM}$
decreases. The DM component increases and can even dominate the NM component as $\sigma_c$ increases 
along the sequences.  
With a large $\rho_c$, the effect of DM is small. This can be seen in the left column of Fig.~\ref{fig:structure}, where $\Delta\bar{Q}/\bar{Q}$ and $M_{\rm DM}$ are relatively small. The 
DM component develops a core structure embedded inside the NM, with both radii decreasing when we increase $\sigma_c$.

With a moderate $\rho_c$ (middle column), the structure of FBSs with small $\sigma_c$ are 
qualitatively the same as those for a large $\rho_c$ described above.  
As we increase the amount of DM admixed, the tidal deformability ${\bar \lambda}_{\rm tidal}$ 
increases until the DM mass becomes comparable to NM mass. Once $M_{\rm NM} < M_{\rm DM}$, we find that the tidal deformability reverses course and begins to decrease instead. At the same time, $M_{\rm NM}$ ($M_{\rm DM}$) continues to decrease (increase). This observation matches the findings in \cite{diedrichs2023tidal}.

On the other hand, $r_{\rm NM}$ also decreases continuously, but $r_{\rm DM}$ is relatively constant.
One significant difference of the stellar structure compared to the scenario with high $\rho_c$ is that the FBSs now have cloud structures. The bosonic field is no longer confined to the core of the star, but extends to a larger radius surrounding the NM.
This specific case also shows that the turning point in the tidal deformability is not caused by 
the transition of the DM from being located at the core of the star to forming a cloud surrounding the NM, as all the FBSs in this sequence exhibit cloud structures. 
Instead, the turning point is likely caused by the crossing at $M_{\rm NM} = M_{\rm DM}$ along the
sequence. When $M_{\rm NM} < M_{\rm DM}$, the trend of tidal deformability changes, which matches with the turning points in all panels. 
We observe that this behavior persists in sequences with other values of $\rho_c$, $m_b$ and $\Lambda$ 
which we do not present here, making us believe that the turning point is directly triggered by the change in mass dominance. 
With a small $\rho_c$, the bosonic part dominates for most configurations and ultimately converges to BS solutions when $\sigma_c$ increases. In the right column of Fig.~\ref{fig:structure}, it is seen that the tidal deformability decreases as $\sigma_c$ increases, which is opposite to the behavior observed 
in the cases of large and moderate $\rho_c$. However, it is worth recalling that there exists a turning point
	in the moderate $\rho_c$ case after which the tidal deformability decreases as $\sigma_c$ increases.      
As the star becomes more massive and dense, the tidal deformability keeps decreasing until it converges to the limit for pure BSs. 
The mass and radius are also consistent with the BS limit. Meanwhile, the tiny fermionic component locates inside the bosonic cloud, which essentially contributes no effect to the stellar structure.

Typical astrophysical observations, both galactic \cite{kiziltan2013the} and gravitational wave \cite{landry2021the} observations, suggest that NSs have masses greater than $1 M_{\odot}$. This mass range roughly corresponds to NSs with $\bar{\lambda}_{\rm tidal} \lesssim 10^3$ for typical nuclear matter EOS models. For more massive NSs up to about $2 M_\odot$, $\bar{\lambda}_{\rm tidal}$ decreases
to $\mathcal{O}(10)$. 
From the first observation of the GW from a binary NS merger GW170817 \cite{abbott2019properties}, 
an upper bound on ${\bar\lambda}_{\rm tidal} < 800$ of a $1.4 M_\odot$ star is obtained. 
It would be interesting to understand the properties of FBSs that violate the I-Love-Q relations, 
and have $M$ and $\bar{\lambda}_{\rm tidal}$ that are relevant in astrophysical studies. 
In Fig.~\ref{fig:ILoveQ}, we find that the relevant configurations are generally NS-like configurations with large $\rho_c$. 
These stars typically are admixed with $5\%-10\%$ DM with total mass $M=1.3-2.0M_{\odot}$. Marginal cases happen at large $\bar{\lambda}_{\rm tidal}$, where stars can admix up to 25\% DM with mass $M\approx0.9M_{\odot}$. 

\begin{figure*}[t]
	\centering
	\includegraphics[width=\textwidth]{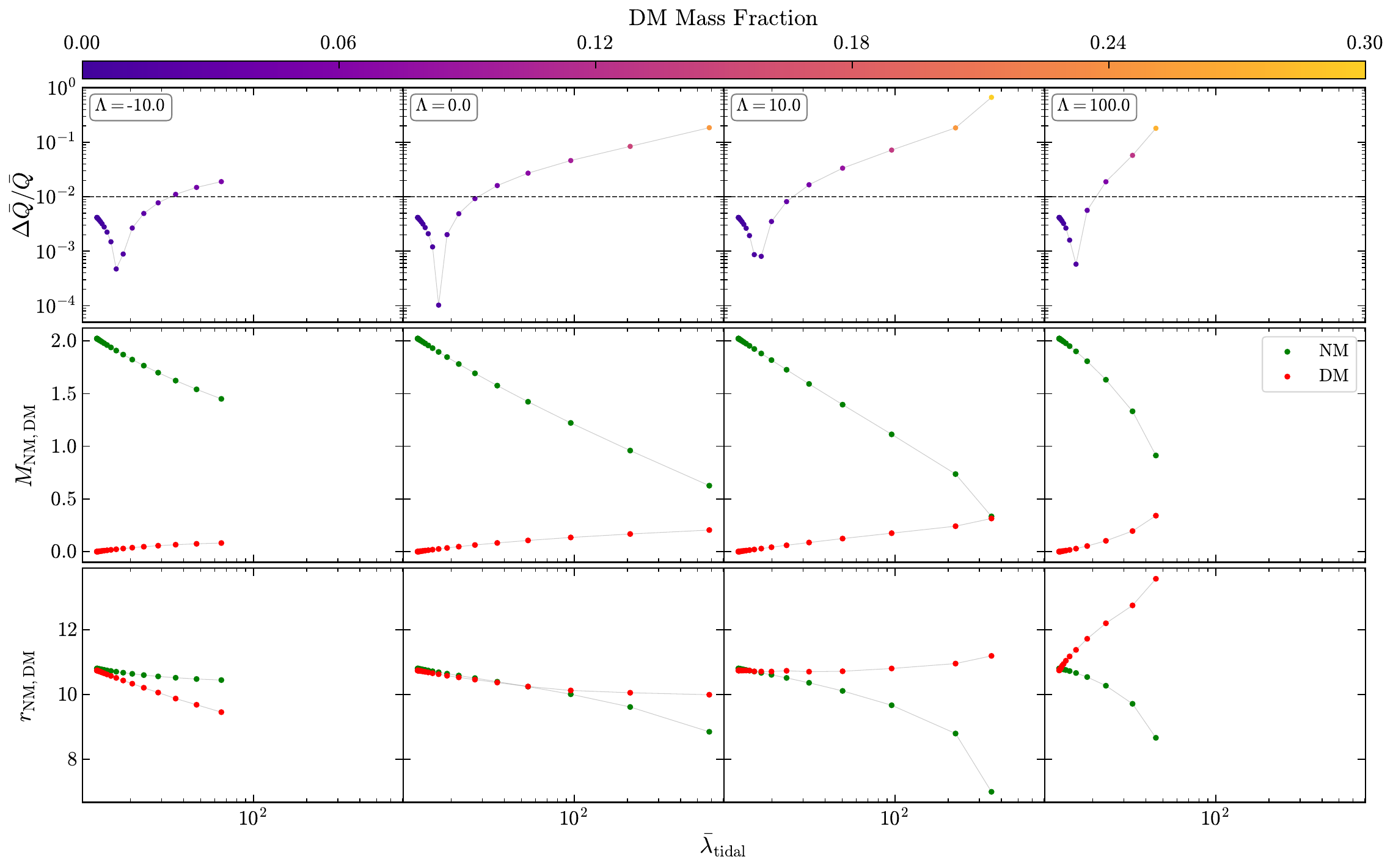}
	\caption{Similar to Fig.~\ref{fig:structure}, but for four sequences of FBSs with 
		$m_b=2.68\times10^{-10}\,\mathrm{eV}$, $\rho_c=1.65\times10^{15}\,\mathrm{g/cm}^3$ and 
		$\Lambda=\{-10,0,10,100\}$ (from left to right). The value of $\rho_c$ is chosen to produce 
		FBSs within the astrophysically relevant range where $\bar{\lambda}_{\rm tidal} \lesssim \mathcal{O}(10^2)$. Note that the colorbar ranges from 0.0 to 0.3, unlike in Fig.~\ref{fig:structure} where it	 ranges from 0.0 to 1.0.	 }
	\label{fig:Q_Love_structure_Lambda}
\end{figure*}

\subsection{Self-interaction strength} \label{subsec:Lambda}

Now we study how the normalized coupling constant $\Lambda$ affects the universal relations. 
In Fig.~\ref{fig:Q_Love_structure_Lambda} we consider four sequences of FBSs with the same 
$m_b=2.68\times 10^{-10}$ eV and $\rho_c=1.65\times 10^{15}\ {\rm g/cm}^3$, but different 
values of $\Lambda =\{-10, 0, 10, 100 \}$ from left to right in the figure. 
Here, we specifically choose the constant value $\rho_c$ so that the resulting values of 
$\bar{\lambda}_{\rm tidal}$ overlap the astrophysical relevant range. 
Similar to Fig.~\ref{fig:structure}, we plot $\Delta {\bar Q}/{\bar Q}$, $M_{\rm NM} (M_{\rm DM})$ and $r_{\rm NM} (r_{\rm DM})$ in each column. We only present the data points for stable FBSs for each 
sequence. The first column presents the results for an attractive self interaction $\Lambda=-10$. 
In this case, the sequence terminates earlier at smaller ${\bar \lambda}_{\rm tidal}$ as we increase
$\sigma_c$ comparing to the sequence without self interaction $\Lambda=0$ (second column), since the attractive interaction causes the stars to become unstable more easily. 
The Q-Love relation can still be satisfied very well in this case. The fractional errors $\Delta {\bar Q}/{\bar Q}$ are consistently smaller than 1\% for the majority of the data points along the sequence. The error increases to about 2\% only towards the end of the sequence. 
Along the sequence, $M_{\rm NM}$ decreases quite significantly from about $2 M_\odot$ to the last stable
configuration at a mass slightly less than $1.5 M_\odot$ as we increase $\sigma_c$. 
On the other hand, the DM mass $M_{\rm DM}$ remains small and increases only slightly. Overall, only
a relatively small amount of attractive DM can exist inside stable FBSs.    
The FBSs in this sequence all have core structures with the NM and DM radii being very similar at the beginning of the sequence. As $\sigma_c$ increases, the DM radius $r_{\rm DM}$ decreases more rapidly than the NM radius $r_{\rm NM}$, resulting in the DM core becoming more compact.

For the sequence without self interaction $\Lambda=0$, the trends of $\Delta {\bar Q}/{\bar Q}$,
$M_{\rm NM}$, and $M_{\rm DM}$ are qualitatively the same as those of the constant 
$\rho_c = 2.03\times 10^{15} \ {\rm g/cm}^3$ sequence in Fig.~\ref{fig:structure}, though the fractional error $\Delta {\bar Q}/{\bar Q}$ can now increase to 10\% level near the end of the sequence. 	
However, with a smaller $\rho_c = 1.65\times 10^{15}\ {\rm g/cm}^3$, the structures of FBSs in this
sequence are quite different. 
Specifically, the DM radius $r_{\rm DM}$ is slightly smaller than the NM radius $r_{\rm NM}$ 
initially, and both decrease as $\sigma_c$ increases along the sequence. 
As $\sigma_c$ continues to increase, the NM radius keeps decreasing when ${\bar \lambda}_{\rm tidal} > 10^2$, while the DM component remains relatively unchanged, forming a cloud structure that surrounds the NM.

Let us now consider the effects of repulsive self interaction with $\Lambda > 0$.  
As we increase $\Lambda$ from 0 to 100 in Fig.~\ref{fig:Q_Love_structure_Lambda}, it is observed 
that the FBS sequence terminates at a smaller ${\bar \lambda}_{\rm tidal}$ as $\sigma_c$ increases. 
The NM (DM) mass also decreases (increases) more rapidly for larger $\Lambda$, resulting in a 
greater change of the stellar structure along the sequence. This leads to a larger fractional 
error $\Delta {\bar Q}/{\bar Q}$ for the same ${\bar \lambda}_{\rm tidal}$.     	
In the bottom panels of Fig.~\ref{fig:Q_Love_structure_Lambda} , we observe a crossing between $r_{\rm NM}$ and $r_{\rm DM}$ occurring at a smaller ${\bar\lambda}_{\rm tidal}$ earlier in the sequence for larger $\Lambda$. Furthermore, the DM radius increases rapidly for larger $\Lambda$. As more DM is admixed, it must extend further away to balance gravity, while simultaneously forcing the fermionic NM matter to form a dense core.

In the top panels of Fig.~\ref{fig:Q_Love_structure_Lambda}, focusing on the points slightly above the 1\% line (up to $\sim$10\%) within the range of astrophysical interest where 
${\bar \lambda}_{\rm tidal} \sim \mathcal{O}(10^2)$, we typically find a DM fraction $f=3\%-20\%$.

The above descriptions about the Q-Love relation are also true for the I-Love relation. 
Increasing $\Lambda$ allows for a larger $M_{\rm DM}$, leading to a greater deviation in $\bar{I}$ for 
a given $\bar{\lambda}_{\rm tidal}$. The changes in the I-Love relation are generally the same as 
those of the Q-Love relation.

\begin{figure}[t]
	\centering
	\includegraphics[width=\columnwidth]{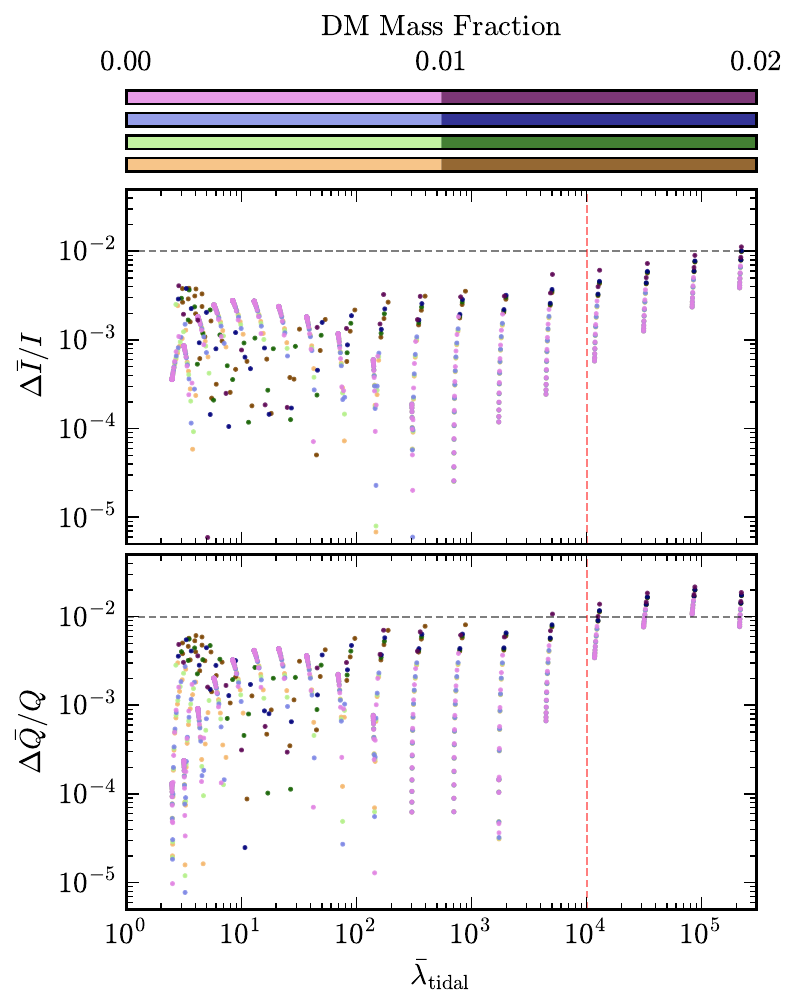}
	\caption{Fractional errors between the I-Love (top) and Q-Love relations (bottom) and FBS data
		for $m_b=2.68\times10^{-10}\,\mathrm{eV}$ and $\Lambda=\{-10, 0, 10, 100\}$. The colorbars represent 
		the DM mass fraction $f$ of configurations, where lighter colors indicate a lower $f$. 
		Colored data points represent stars with different values of $\Lambda$:
		$\Lambda = -10$ (brown), 0 (green), 10 (blue), and 100 (purple). 
	}
	\label{fig:I_Love_Q_error_all}
\end{figure}

We also observe that the maximum value of $\Delta\bar{Q}/\bar{Q}$ depends significantly on $f$ rather than on 
$\Lambda$. With the same $f$, the maximum fractional error of $\bar{Q}$ is almost the same for all $\Lambda$ values\footnote{Physically, the increase of $\Delta \bar{Q}/\bar{Q}$ depends greatly on the increase of $M_{\rm DM}$. As the star contains more DM, the non-rotating DM component causes a deviation in $\bar{Q}$ from the I-Love-Q relation. The deviations of $\bar{Q}$ and $\bar{I}$ are sensitive to the DM mass fraction $f$. On the other hand, the insensitivity to $\Lambda$ can be explained by the fact that the maximum fractional errors of $\bar I$ and $\bar Q$ are always attained by stars with a tiny scalar field. This suppresses the self-interaction term (see Eq.~(\ref{eq:FBS})), which leads to a subleading effect compared to that of $f$. }.
Hence, an interesting question is whether there is an upper limit for $f$, independent of $\Lambda$, such that the 
I-Love-Q relations remain universal? 
To address this question, we plot the errors of the I-Love and Q-Love relations for configurations with $f\leq0.02$ in Fig.~\ref{fig:I_Love_Q_error_all}. 
We use different colors to represent the data points for different values of $\Lambda$: 
$\Lambda=-10$ (brown), 0 (green), 10 (blue), and 100 (purple).   
In the top panel of Fig.~\ref{fig:I_Love_Q_error_all}, we see that the I-Love relation lies well within the 1\% error bar. The I-Love relation remains a universal relation for FBSs with 2\% DM admixed. Even for $f\leq0.04$, we observe that the I-Love fitting curve for pure NSs is still valid. 
On the other hand, the Q-Love relation gives a more stringent limit, in which configurations with $f\gtrsim0.02$ would violate the universal relation. 
Note, however, that this conclusion is true only for $m_b\geq 2.68\times10^{-10} \ \mathrm{eV}$ (see Sec.~\ref{subsec:boson_mass}). 	

\subsection{Boson particle mass} \label{subsec:boson_mass}

\begin{figure*}[t]
	\includegraphics[width=\textwidth]{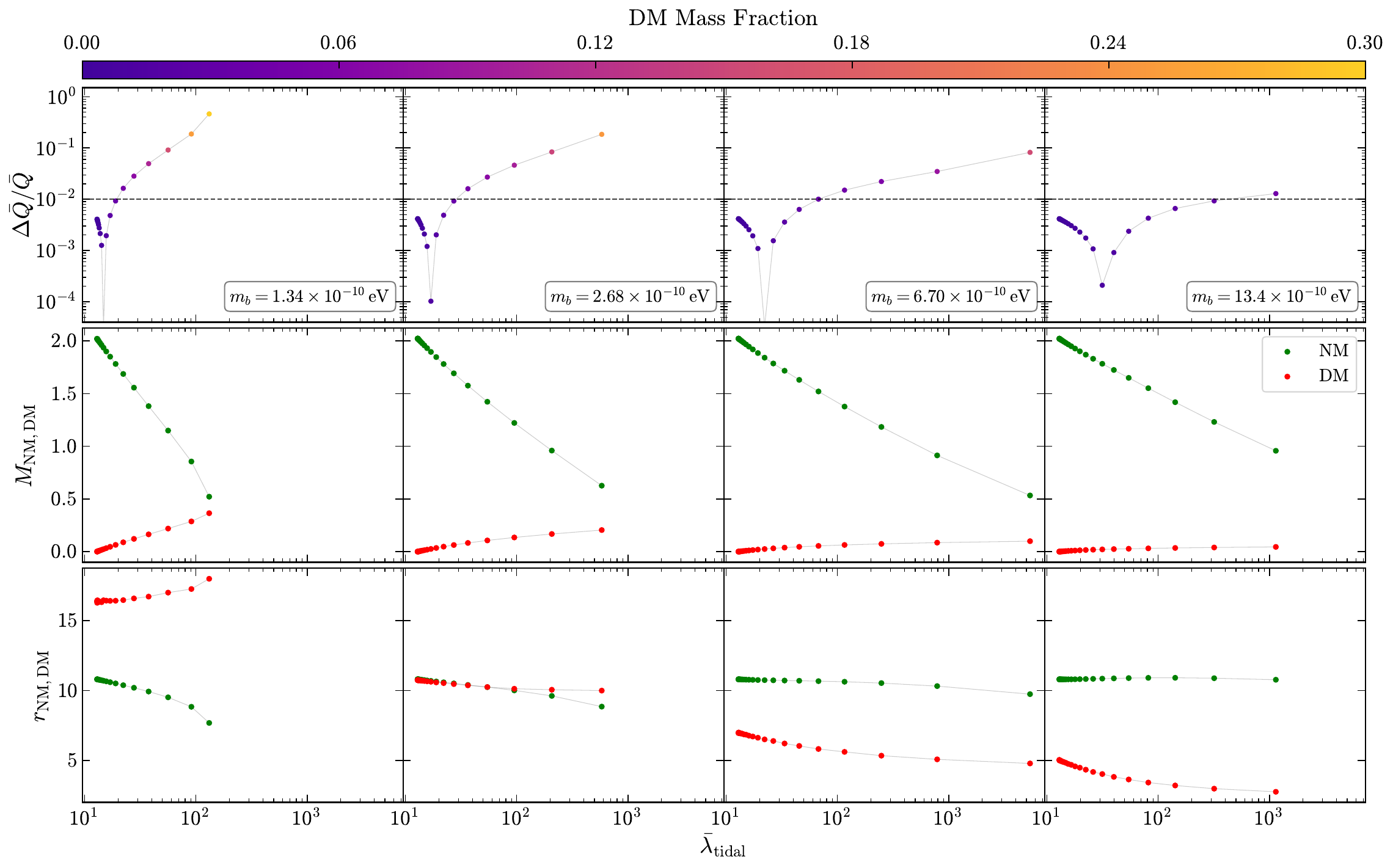}
	\caption{
		Similar to Fig.~\ref{fig:structure}, but for four sequences of FBSs with $\Lambda=0$, 
		$\rho_c=1.65\times10^{15}\,\mathrm{g/cm}^3$ and $m_b=\{1.34,2.68,6.70,13.4\}\times10^{-10}\,\mathrm{eV}$
		(from left to right). Note that the colorbar ranges from 0.0 to 0.3, unlike in Fig.~\ref{fig:structure}
		where it ranges from 0.0 to 1.0. }
	\label{fig:Q_Love_structure_m}
\end{figure*}

We further investigate the effect of bosonic particle mass. We have tested four different boson particle masses, $m_b=\{1.34,2.68,6.70,13.4\}\times10^{-10}\,\mathrm{eV}$. These particle masses fall within a range where the boson's Compton wavelength is roughly equal to the Schwarzschild radius of a 
$1M_{\odot}$ star. In Fig.~\ref{fig:Q_Love_structure_m}, we present the results for these four 
sequences of FBSs with different $m_b$, all having the same values of $\Lambda = 0$ and 
$\rho_c = 1.65 \times 10^{15}\ {\rm g/cm}^3$. 	
We take the same value of $\rho_c$ as in Fig.~\ref{fig:Q_Love_structure_Lambda} to illustrate the 
results within the relevant astrophysical range. 
First, we see that the fractional error $\Delta {\bar Q}/{\bar Q}$ is larger for lighter DM particles. These FBSs are restricted to have a relatively lower $\lambda_{\rm tidal}$, placing them within 
our range of interest.
In the middle row of Fig.~\ref{fig:Q_Love_structure_m}, which displays the trends of $M_{\rm NM}$ and $M_{\rm DM}$, it is observed that a smaller $m_b$ leads to a larger $M_{\rm DM}$, resulting in a greater deviation of the FBS data from the Q-Love relation for pure NSs. 
As we increase the particle mass to $m_b=13.4\times10^{-10}\,\mathrm{eV}$, the fractional errors in both $\bar{Q}$ and $\bar{I}$ for all FBSs typically decrease to within 1\% level. 
Following this trend, we expect that the universal relations will remain valid for even larger values of $m_b$.

In the bottom row of Fig.~\ref{fig:Q_Love_structure_m}, it is observed that the bosonic DM radii 
$r_{\rm DM}$ of FBSs decrease as $m_b$ increases, and coincidentally, the chosen boson mass range
leads to a structural change in FBSs, as described in~\cite{diedrichs2023tidal}. For 
the sequence with the smallest boson mass $m_b=1.34\times 10^{-10}$ eV, the bosonic components develop
cloud structures that enclose the NM cores. FBSs with any smaller boson particle mass would result in more extended cloud structures. Along this sequence, $r_{\rm DM}$ also increases with 
the value of $\sigma_c$. For $m_b=2.68 \times 10^{-10}$ eV, we have $r_{\rm DM} \approx r_{\rm NM}$, 
except near the end of the sequence where $r_{\rm DM}$ becomes slightly larger than $r_{\rm NM}$. 
FBSs with more massive bosons ($m_b > 2.68 \times 10^{-10}$ eV) form a DM core inside, where 
the radius $r_{\rm DM}$ decreases as $\sigma_c$ increases along the sequence.

For the case of FBSs with a DM cloud structure, electromagnetic observations of such a FBS would give  	
the surface radius of the fermionic NM component $r_{\rm NM}$, while measurements sensitive only 
to gravitational effects would give the exterior surface radius $r_s (= r_{\rm DM})$. 
In some parts of the boson parameter space, such as the sequence for 
$m_b=1.34\times 10^{-10}$ eV, these stars can easily deviate from the I-Love-Q relations by a significant amount, and hence 
independent measurements of the relevant quantities of the universal relations can in principle be used to test their existence. On the other hand, for FBSs with a DM core, both electromagnetic and gravitational observations would measure the same radius, namely the fermionic NM radius $r_{\rm NM}$.

In Sec.~\ref{subsec:Lambda}, we focused on $m_b=2.68\times 10^{-10} \ \mathrm{eV}$ and
determined an upper bound for the DM fraction $f=0.02$ above which the I-Love-Q relations are violated by more than the 1\% level. Extending the analysis to other values of $m_b$, we found that the upper bound $f=0.02$ still holds for FBSs with $m_b\geq2.68\times10^{-10} \ \mathrm{eV}$. For $m_b=\{6.70,13.4\}\times10^{-10} \ \mathrm{eV}$, however, the upper bound can increase to $f\approx0.03$. As for $m_b=1.34\times10^{-10}\,\mathrm{eV}$, in order to stay within the 1\% level of the I-Love-Q relation, 
$f\leq 0.01$ is needed. As FBSs with any smaller $m_b$ will have an even heavier bosonic component, it is expected that any FBSs with $m_b<1.34\times10^{-10}\,\mathrm{eV}$ will need an even lower $f$ to satisfy the I-Love-Q relations. 
Therefore, the upper bound for $f$ in fact depends on $m_b$, though it is insensitive to $\Lambda$ for a fixed value of $m_b$.

When considering different self-interaction strengths $\Lambda=\{-10,0,10,100\}$, we empirically find
that with $m_b=26.8\times10^{-10}\,\mathrm{eV}$, all stable configurations within the valid range of
I-Love-Q relations have $\Delta\bar{Q}/\bar{Q}<1\%$ as the mass of the bosonic component of the star becomes insignificant. This effectively sets an upper bound on the boson particle mass of FBSs that can be explored by investigating the I-Love-Q relation violations.

As a side note before ending this section, recall that electromagnetic observations would give the surface radius of the fermionic component $r_{\rm NM}$, while measurements sensitive only to gravity give the exterior surface radius $r_s$. The two radii are the same for FBSs with a bosonic core structure. 
However, they are different ($r_{\rm NM} < r_s$) for FBSs with a cloud structure, and hence simultaneous 
electromagnetic and gravitational measurements could in principle be used to probe the cloud structure. 
As seen from the lower panels of Fig.~\ref{fig:Q_Love_structure_m}, we notice that core structures only 
happen when $m_b>1.34\times10^{-10}\,\mathrm{eV}$. 
Incorporating this constraint into consideration, one can place a bound 
$m_b \leq 1.34\times10^{-10}\,\mathrm{eV}$ if a FBS with cloud structure is observed.

\section{Equation-of-State Dependence} \label{sec:eos}

Although the original I-Love-Q relations are approximately insensitive to the choice of EOS, the deviation due to DM could depend sensitively on EOS. 
Specifically, using an EOS with a vastly different mass-radius relation, comparing to the APR EOS employed in Sec.~\ref{sec:result}, can change the structure of the fermionic component and affect the values of the 
I-Love-Q trio. The bounds on the boson mass range discussed in the previous section might thus be altered. 
Therefore, it is necessary to examine a different EOS to determine whether our previous conclusions remain valid.

\begin{figure}[tb]
	\includegraphics[width=\columnwidth]{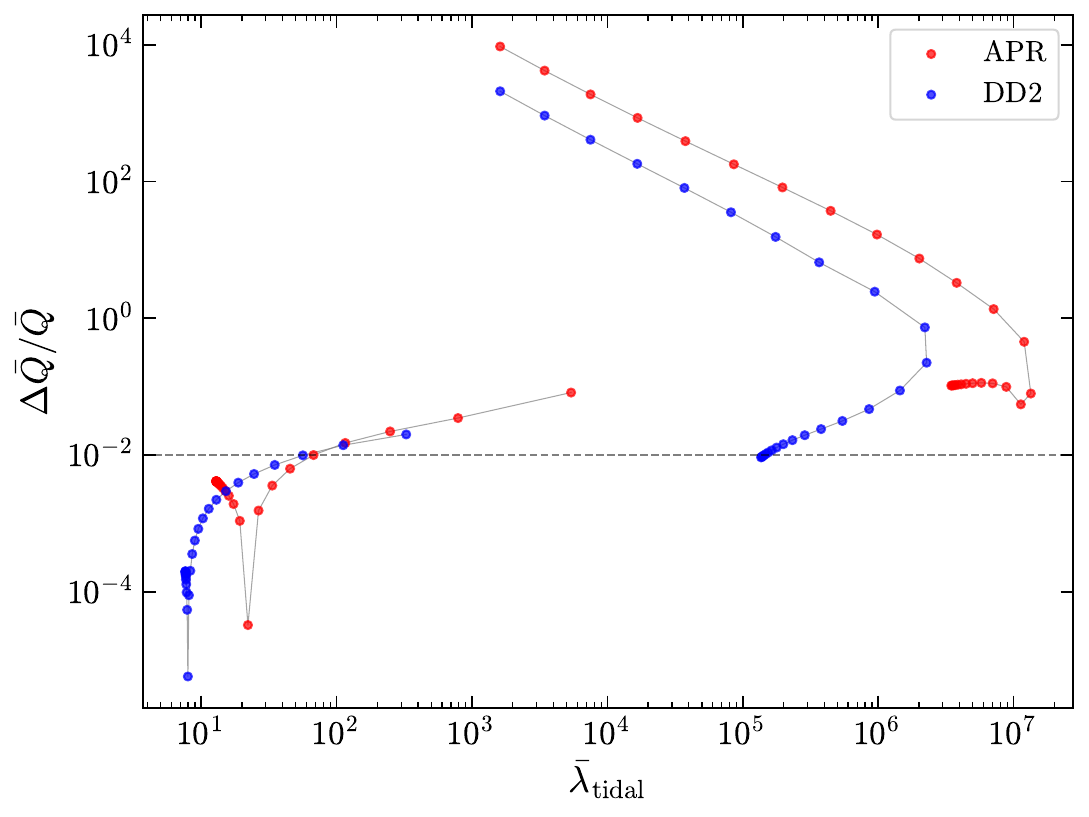}
	\caption{ 
	Comparison between the fractional errors $\Delta {\bar Q}/{\bar Q}$ for FBSs computed with the APR (red lines) and DD2 EOS (blue lines). Two different values of $\rho_c$ are taken to illustrate the EOS dependence in different $\bar{\lambda}_{\rm tidal}$ regimes. The two sequences on the right (left) are obtained by fixing $\rho_c=0.35 \times 10^{15} \ \mathrm{g/cm}^3$ ($1.65\times 10^{15}\ \mathrm{g/cm}^3$).  
	The boson particle mass and self-interaction strength are fixed at $m_b=2.68\times10^{-10} \ \mathrm{eV}$ and $\Lambda=0$, respectively.  
	}
	\label{fig:EOS_dependence}
\end{figure}

We have chosen the DD2 EOS \cite{hempel2010a, typel2010composition} to explore the dependence on EOS,
as its mass-radius relation significantly differs from that produced by the APR EOS. 
For a given stellar mass, NSs constructed by the DD2 EOS typically give a larger radius compared to
those constructed using other commonly employed EOSs. 
By choosing this EOS, we aim to demonstrate how the deviations of universal relations may change. 	
In Fig.~\ref{fig:EOS_dependence}, we show the fractional error $\Delta\bar{Q}/\bar{Q}$ for two sets of sequences corresponding to the APR (red) and DD2 (blue) EOS. All sequences are computed with 
$m_b=2.68\times10^{-10} \ \mathrm{eV}$ and $\Lambda=0$. The sequences on the left are chosen to have the same NM central density $\rho_c=1.65\times10^{15}\,\mathrm{g/cm}^3$ as in Figs.~\ref{fig:Q_Love_structure_Lambda}~and~\ref{fig:Q_Love_structure_m}, which overlaps with the relevant astrophysical range. 
The sequences on the right have $\rho_c=0.35\times10^{15}\,\mathrm{g/cm}^3$, the same as the right column in Fig.~\ref{fig:structure}, to illustrate the behavior at lower $\rho_c$ values.

On the left, fixing the central densities, configurations that are computed with the APR EOS have a larger tidal deformability when compared to those computed with the DD2 EOS. The fractional errors $\Delta\bar{Q}/\bar{Q}$ for both sequences lie in the same range. Given the same $\Delta\bar{Q}/\bar{Q}$, FBSs on both sequences have a similar DM mass fraction. This suggests that the deviation remains insensitive to the choice of EOS in this regime.

On the right, most APR configurations have a larger tidal deformability, except those at the end of the sequences. Those stars have a tiny fermionic component that has little impact on the tidal deformabilities,
resulting in similar values. 	However, the difference in EOS leads to differences in NM structure within the stars, resulting in noticeable discrepancies in $\Delta\bar{Q}/\bar{Q}$. This effect is similar to that in Fig.~\ref{fig:ILoveQ}, where sequences remain parallel in the bosonic limit.

We now aim to test the validity of the bounds established in previous sections.
First, we replicated Fig.~\ref{fig:I_Love_Q_error_all} with the DD2 EOS in order to test the upper limit on 
the DM mass fraction $f$. We find that $\Delta\bar{Q}/\bar{Q}$ remains relatively insensitive to $\Lambda$, 
and the upper bound on $f$ remains true for $m_b\geq2.68\times10^{-10}\,\mathrm{eV}$. 
In other words, the I-Love-Q relations are still satisfied to high accuracy and cannot be used to identify 
FBSs with a DM admixture of 2\% or less. 
However, for FBSs with heavier boson particle mass such as $m_b=6.70\times10^{-10} \ \mathrm{eV}$, the upper limit stated in Sec.~\ref{subsec:boson_mass} no longer holds. Instead of a 3\% limit, it is suppressed to 2\%. This shows that the deviation due to DM depends slightly on the EOS, which could potentially be removed by sampling over a class of commonly used EOS.

Furthermore, using the DD2 EOS, $m_b\geq26.8\times10^{-10} \ \mathrm{eV}$ remains to be an upper bound
in the sense that all stable configurations within the valid range of I-Love-Q relations have 
$\Delta\bar{Q}/{\bar{Q}} < 1\%$ when considering different self-interaction strengths, and thus boson
particle mass above this bound cannot be explored by investigating the I-Love-Q relation violations. 	 
Since this upper bound is strongly linked to the fact that large $m_b$ leads to a small DM mass, leading
to a negligible impact on the values of the I-Love-Q trio, we believe that the upper bound for $m_b$   
remains valid regardless of the EOS employed.

\begin{table*}[ht]
\centering
\def\arraystretch{1.5}
\setlength{\tabcolsep}{3.5pt}
\begin{tabular}{c@{\hskip 12pt}ccccccccccc}
\toprule
\multicolumn{1}{l}{} & \multicolumn{4}{c}{Input Parameters} & & \multicolumn{6}{c}{Global Quantities} \\ \cmidrule(){2-5} \cmidrule(){7-12}
Ref. & $\rho_c$ & $\sigma_c$ & $m_b$ & $\Lambda$ & & $\bar{\lambda}_{\rm tidal}$ & $M_{\rm NM}$ & $M_{\rm DM}$ & $r_{\rm NM}$ & $r_{\rm DM}$ & $\Delta \bar{Q} / \bar{Q}$ \\ \midrule
Fig. \ref{fig:structure} & ↑ & Fixed & Fixed & 0 & & ↓ & ↑ & ↓ & ↑ & ↓ & ↓ \\
Fig. \ref{fig:structure} & Fixed & ↑ & Fixed & 0 & & Depends on $\rho_c$ and $\sigma_c$ & ↓ & ↑ & ↓ & ↓ & ↑ \\
Fig. \ref{fig:Q_Love_structure_Lambda} & Fixed & Fixed & Fixed & ↑ & & $\sim$ & ↓ & ↑ & ↓ & ↑ & $\sim$ \\
Fig. \ref{fig:Q_Love_structure_Lambda} & Fixed & ↑ & Fixed & Fixed & & ↑ & ↓ & ↑ & ↓ & ↑ & $\sim$ \\
Fig. \ref{fig:Q_Love_structure_m} & Fixed & Fixed & ↑ & 0 & & ↑ & ↑ & ↓ & $\sim$ & ↓ & ↓ \\
Fig. \ref{fig:Q_Love_structure_m} & Fixed & ↑ & Fixed & 0 & & ↑ & ↓ & ↑ & Depends on $m_b$ & Depends on $m_b$ & ↑ \\ \bottomrule
\end{tabular}
\caption{Summary of the general trend for global quantities of FBSs. By altering one of the input parameters, we
list the behaviors of various global quantities in response to the change.
Symbols ↑ , ↓ and $\sim$ represent that the quantity increases, decreases and barely changes, respectively.}
\label{tab:summary}
\end{table*}

\section{Conclusion} \label{sec:conclusion}

In this work, we have studied the I-Love-Q relations for a class of DM-admixed NSs known as FBSs, 
where the DM component is modeled as a complex scalar bosonic field. The tidal deformability of 
nonrotating FBSs is determined by solving the linearized Einstein-Klein-Gordon equations \cite{diedrichs2023tidal}. 	
For the computation of the moment of inertia and spin-induced quadrupole moment, we extend the 
standard formulation for pure NSs \cite{hartle1967slowly} to FBSs comprising a slowly rotating 
fermionic NM component admixed with nonrotating bosonic DM. 
Depending on the model parameters, we found that the bosonic component can form either a compact core 
enclosed within the NM or a cloud-like structure that extends beyond the NM radius, agreeing with
findings from previous investigations \cite{diedrichs2023tidal}. 
In stellar models with a bosonic core, the fermionic NM dominates and the stellar structures
are similar to those of pure NSs. As a result, the I-Love-Q relations can still be satisfied to within 
a few percent level.
On the other hand, FBSs with a bosonic cloud-like structure generally can lead to larger deviations of the
I-Love-Q relations. 
When the mass fraction of DM approaches unity, the properties of FBSs, such as the maximum mass 
and tidal deformability, converge towards those of pure BSs in this limit. 

We also examined the effect of self-interaction strength $\Lambda$ and boson particle mass $m_b$. For FBSs with the same central densities and $m_b$, a smaller $\Lambda$ leads to a smaller DM mass and larger NM mass. An attractive self interaction causes the DM to form a dense core, while a repulsive self interaction causes the DM to extend further away to form a cloud-like structure. 
On the other hand, varying $m_b$ while fixing $\Lambda$ also changes the structure of FBSs. A smaller $m_b$ allows FBSs to admix more DM, which extends to form cloud structures. Larger $m_b$ lowers the DM mass and forces a DM core to form, leading to a smaller deviation of the I-Love-Q relations. 
We summarize the main trends of the numerical results in Table \ref{tab:summary}, which illustrates how different global quantities change when we alter the input parameters. 
For instance, the first row of data in the table summarizes the results in Fig.~\ref{fig:structure} when 
comparing different values of $\rho_c$ for the same (fixed) $\sigma_c$. The second row of data summarizes the
trends in Fig.~\ref{fig:structure} when $\sigma_c$ increases along a sequence of fixed $\rho_c$.
Specifically, the trend of ${\bar \lambda}_{\rm tidal}$ depends on the chosen value of $\rho_c$ and also the value of $\sigma_c$ along the sequence.

By taking a slice of FBSs of constant $\Delta\bar{Q}/\bar{Q}$, we found that the deviation is highly correlated with DM mass fraction $f$, but fairly independent of $\Lambda$. For a given boson particle mass $m_b\geq2.68\times10^{-10} \ \mathrm{eV}$, FBSs with DM mass fraction $f<0.02$ 
computed by the APR EOS could not be discerned by the I-Love-Q relations. As we increase (decrease) $m_b$, the upper limit increases (decreases). When the particle mass increases further to $m_b\geq26.8\times10^{-10} \ \mathrm{eV}$, the DM contribution becomes insignificant and the deviation of the I-Love-Q relations is bound within the EOS-sensitivity level (about 1\%) of the universal relations, making these FBSs indistinguishable from NSs according to the I-Love-Q relations. This sets an upper bound for the boson mass range that can be explored by investigating the I-Love-Q relation violations.

Although the original I-Love-Q relations are insensitive to the EOS choice, the deviation due to DM still depends on the selected EOS. 
In order to compare the results obtained by the APR EOS as summarized above, we have also employed the
DD2 EOS in our study and found that the upper limit of $f$ changes with the choice of EOS, verifying that the EOS dependence in the deviations of I-Love-Q relations cannot be ignored. 
This issue could be addressed by incorporating more EOS into the analysis in future investigation.  
On the other hand, the upper bound of the boson particle mass $m_b=26.8\times 10^{-10}$ eV does not change. Since this upper bound is strongly related to the DM mass confined in FBSs, but not the structure of fermionic NM component, we expect this upper bound to be insensitive to the NM EOS.

We end this paper with a few remarks. (1) While we have studied how the deviations of the I-Love-Q relations 
depend on the properties of DM components of FBSs, it remains challenging to simultaneously measure the relevant physical quantities from a single compact star to test the predicted deviations.
A more feasible approach may use the fact that universal relations can help reduce the number of matter parameters in theoretical gravitational waveform modelling for  binary NS inspirals \cite{Lackey:2019,Schmidt:2019,Barkett:2020,Andersson:2021}. 
The effects of DM may then be explored by comparing waveform models with and without the assumption of universal relations against observational data.   
(2) In this work, we assume general relativity is the correct theory of gravity. Otherwise, any observed deviations 
of the I-Love-Q relations may also be due to the effects of modified gravity theories \cite{Sham:2014,berti2015testing,pani2015iloveq,doneva2015iq,gupta2018iloveq} instead of DM alone. 
(3) Even within the theory of general relativity, it should be noted that there are known effects of NS physics that can break the universal relations, such as slowly rotating NSs with very strong magnetic field \cite{Haskell:2013} and thermal effects in newborn NSs \cite{Martinon:2014,Marques:2017}. 
(4) Finally, it would be interesting to extend our study to FBSs with rotating bosonic components. 
As discussed in Sec.~\ref{subsec:I}, slowly rotating pure BSs cannot be constructed by performing perturbative calculations about the nonrotating background as the assumption of axisymmetry imposes a boundary condition on the scalar field, leading to the quantization of angular momentum. Similar consideration should also apply to rotating FBSs. 
Nevertheless, in principle, one should be able to extend the calculation of rotating BSs to generic rotating FBSs by coupling the hydrodynamics equations for the fermionic component with the Einstein-Klein-Gordon system without the need for the approximation of slow rotation. We leave this issue for future investigation.
On the other hand, one can bypass this technical issue by focusing only on the limit of strong 
self-interaction ($\Lambda \gg 1$) for the bosonic component. In this regime, the scalar field can be 
effectively modeled as a perfect fluid with a specific EOS \cite{chavanis2012bose}. 
A slowly rotating FBS can then be modeled as a two-fluid system, similar to the study of superfluid NSs \cite{Yeung:2021}.

\section*{Acknowledgment}
This work is supported by a grant from the Research Grants Council of the Hong Kong SAR, China (Project No: 14304322).	


\begin{thebibliography}{71}%
	\makeatletter
	\providecommand \@ifxundefined [1]{%
	 \@ifx{#1\undefined}
	}%
	\providecommand \@ifnum [1]{%
	 \ifnum #1\expandafter \@firstoftwo
	 \else \expandafter \@secondoftwo
	 \fi
	}%
	\providecommand \@ifx [1]{%
	 \ifx #1\expandafter \@firstoftwo
	 \else \expandafter \@secondoftwo
	 \fi
	}%
	\providecommand \natexlab [1]{#1}%
	\providecommand \enquote  [1]{``#1''}%
	\providecommand \bibnamefont  [1]{#1}%
	\providecommand \bibfnamefont [1]{#1}%
	\providecommand \citenamefont [1]{#1}%
	\providecommand \href@noop [0]{\@secondoftwo}%
	\providecommand \href [0]{\begingroup \@sanitize@url \@href}%
	\providecommand \@href[1]{\@@startlink{#1}\@@href}%
	\providecommand \@@href[1]{\endgroup#1\@@endlink}%
	\providecommand \@sanitize@url [0]{\catcode `\\12\catcode `\$12\catcode `\&12\catcode `\#12\catcode `\^12\catcode `\_12\catcode `\%12\relax}%
	\providecommand \@@startlink[1]{}%
	\providecommand \@@endlink[0]{}%
	\providecommand \url  [0]{\begingroup\@sanitize@url \@url }%
	\providecommand \@url [1]{\endgroup\@href {#1}{\urlprefix }}%
	\providecommand \urlprefix  [0]{URL }%
	\providecommand \Eprint [0]{\href }%
	\providecommand \doibase [0]{https://doi.org/}%
	\providecommand \selectlanguage [0]{\@gobble}%
	\providecommand \bibinfo  [0]{\@secondoftwo}%
	\providecommand \bibfield  [0]{\@secondoftwo}%
	\providecommand \translation [1]{[#1]}%
	\providecommand \BibitemOpen [0]{}%
	\providecommand \bibitemStop [0]{}%
	\providecommand \bibitemNoStop [0]{.\EOS\space}%
	\providecommand \EOS [0]{\spacefactor3000\relax}%
	\providecommand \BibitemShut  [1]{\csname bibitem#1\endcsname}%
	\let\auto@bib@innerbib\@empty
	\bibitem [{\citenamefont {{Chatziioannou}}\ \emph {et~al.}(2015)\citenamefont {{Chatziioannou}}, \citenamefont {{Yagi}}, \citenamefont {{Klein}}, \citenamefont {{Cornish}},\ and\ \citenamefont {{Yunes}}}]{chatziioannou2015probing}%
	  \BibitemOpen
	  \bibfield  {author} {\bibinfo {author} {\bibfnamefont {K.}~\bibnamefont {{Chatziioannou}}}, \bibinfo {author} {\bibfnamefont {K.}~\bibnamefont {{Yagi}}}, \bibinfo {author} {\bibfnamefont {A.}~\bibnamefont {{Klein}}}, \bibinfo {author} {\bibfnamefont {N.}~\bibnamefont {{Cornish}}},\ and\ \bibinfo {author} {\bibfnamefont {N.}~\bibnamefont {{Yunes}}},\ }\bibfield  {title} {\bibinfo {title} {{Probing the internal composition of neutron stars with gravitational waves}},\ }\href {https://doi.org/10.1103/PhysRevD.92.104008} {\bibfield  {journal} {\bibinfo  {journal} {\prd}\ }\textbf {\bibinfo {volume} {92}},\ \bibinfo {eid} {104008} (\bibinfo {year} {2015})}\BibitemShut {NoStop}%
	\bibitem [{\citenamefont {{Burgio}}\ \emph {et~al.}(2021)\citenamefont {{Burgio}}, \citenamefont {{Schulze}}, \citenamefont {{Vida{\~n}a}},\ and\ \citenamefont {{Wei}}}]{burgio2021neutron}%
	  \BibitemOpen
	  \bibfield  {author} {\bibinfo {author} {\bibfnamefont {G.~F.}\ \bibnamefont {{Burgio}}}, \bibinfo {author} {\bibfnamefont {H.~J.}\ \bibnamefont {{Schulze}}}, \bibinfo {author} {\bibfnamefont {I.}~\bibnamefont {{Vida{\~n}a}}},\ and\ \bibinfo {author} {\bibfnamefont {J.~B.}\ \bibnamefont {{Wei}}},\ }\bibfield  {title} {\bibinfo {title} {{Neutron stars and the nuclear equation of state}},\ }\href {https://doi.org/10.1016/j.ppnp.2021.103879} {\bibfield  {journal} {\bibinfo  {journal} {Progress in Particle and Nuclear Physics}\ }\textbf {\bibinfo {volume} {120}},\ \bibinfo {pages} {103879} (\bibinfo {year} {2021})}\BibitemShut {NoStop}%
	\bibitem [{\citenamefont {{Persic}}\ \emph {et~al.}(1996)\citenamefont {{Persic}}, \citenamefont {{Salucci}},\ and\ \citenamefont {{Stel}}}]{persic1996the}%
	  \BibitemOpen
	  \bibfield  {author} {\bibinfo {author} {\bibfnamefont {M.}~\bibnamefont {{Persic}}}, \bibinfo {author} {\bibfnamefont {P.}~\bibnamefont {{Salucci}}},\ and\ \bibinfo {author} {\bibfnamefont {F.}~\bibnamefont {{Stel}}},\ }\bibfield  {title} {\bibinfo {title} {{The universal rotation curve of spiral galaxies {\textemdash} I. The dark matter connection}},\ }\href {https://doi.org/10.1093/mnras/278.1.27} {\bibfield  {journal} {\bibinfo  {journal} {MNRAS}\ }\textbf {\bibinfo {volume} {281}},\ \bibinfo {pages} {27} (\bibinfo {year} {1996})}\BibitemShut {NoStop}%
	\bibitem [{\citenamefont {{de Martino}}\ \emph {et~al.}(2020)\citenamefont {{de Martino}}, \citenamefont {{Chakrabarty}}, \citenamefont {{Cesare}}, \citenamefont {{Gallo}}, \citenamefont {{Ostorero}},\ and\ \citenamefont {{Diaferio}}}]{demartino2020dark}%
	  \BibitemOpen
	  \bibfield  {author} {\bibinfo {author} {\bibfnamefont {I.}~\bibnamefont {{de Martino}}}, \bibinfo {author} {\bibfnamefont {S.~S.}\ \bibnamefont {{Chakrabarty}}}, \bibinfo {author} {\bibfnamefont {V.}~\bibnamefont {{Cesare}}}, \bibinfo {author} {\bibfnamefont {A.}~\bibnamefont {{Gallo}}}, \bibinfo {author} {\bibfnamefont {L.}~\bibnamefont {{Ostorero}}},\ and\ \bibinfo {author} {\bibfnamefont {A.}~\bibnamefont {{Diaferio}}},\ }\bibfield  {title} {\bibinfo {title} {{Dark Matters on the Scale of Galaxies}},\ }\href {https://doi.org/10.3390/universe6080107} {\bibfield  {journal} {\bibinfo  {journal} {Universe}\ }\textbf {\bibinfo {volume} {6}},\ \bibinfo {eid} {107} (\bibinfo {year} {2020})}\BibitemShut {NoStop}%
	\bibitem [{\citenamefont {{Padmanabhan}}\ and\ \citenamefont {{Finkbeiner}}(2005)}]{padmanabhan2005detecting}%
	  \BibitemOpen
	  \bibfield  {author} {\bibinfo {author} {\bibfnamefont {N.}~\bibnamefont {{Padmanabhan}}}\ and\ \bibinfo {author} {\bibfnamefont {D.~P.}\ \bibnamefont {{Finkbeiner}}},\ }\bibfield  {title} {\bibinfo {title} {{Detecting dark matter annihilation with CMB polarization: Signatures and experimental prospects}},\ }\href {https://doi.org/10.1103/PhysRevD.72.023508} {\bibfield  {journal} {\bibinfo  {journal} {\prd}\ }\textbf {\bibinfo {volume} {72}},\ \bibinfo {eid} {023508} (\bibinfo {year} {2005})}\BibitemShut {NoStop}%
	\bibitem [{\citenamefont {{Wittman}}\ \emph {et~al.}(2000)\citenamefont {{Wittman}}, \citenamefont {{Tyson}}, \citenamefont {{Kirkman}}, \citenamefont {{Dell'Antonio}},\ and\ \citenamefont {{Bernstein}}}]{wittman2000detection}%
	  \BibitemOpen
	  \bibfield  {author} {\bibinfo {author} {\bibfnamefont {D.~M.}\ \bibnamefont {{Wittman}}}, \bibinfo {author} {\bibfnamefont {J.~A.}\ \bibnamefont {{Tyson}}}, \bibinfo {author} {\bibfnamefont {D.}~\bibnamefont {{Kirkman}}}, \bibinfo {author} {\bibfnamefont {I.}~\bibnamefont {{Dell'Antonio}}},\ and\ \bibinfo {author} {\bibfnamefont {G.}~\bibnamefont {{Bernstein}}},\ }\bibfield  {title} {\bibinfo {title} {{Detection of weak gravitational lensing distortions of distant galaxies by cosmic dark matter at large scales}},\ }\href {https://doi.org/10.1038/35012001} {\bibfield  {journal} {\bibinfo  {journal} {\nat}\ }\textbf {\bibinfo {volume} {405}},\ \bibinfo {pages} {143} (\bibinfo {year} {2000})}\BibitemShut {NoStop}%
	\bibitem [{\citenamefont {{Duffy}}\ and\ \citenamefont {{van Bibber}}(2009)}]{duffy2009axions}%
	  \BibitemOpen
	  \bibfield  {author} {\bibinfo {author} {\bibfnamefont {L.~D.}\ \bibnamefont {{Duffy}}}\ and\ \bibinfo {author} {\bibfnamefont {K.}~\bibnamefont {{van Bibber}}},\ }\bibfield  {title} {\bibinfo {title} {{Axions as dark matter particles}},\ }\href {https://doi.org/10.1088/1367-2630/11/10/105008} {\bibfield  {journal} {\bibinfo  {journal} {New Journal of Physics}\ }\textbf {\bibinfo {volume} {11}},\ \bibinfo {eid} {105008} (\bibinfo {year} {2009})}\BibitemShut {NoStop}%
	\bibitem [{\citenamefont {{Choi}}\ \emph {et~al.}(2021)\citenamefont {{Choi}}, \citenamefont {{Im}},\ and\ \citenamefont {{Shin}}}]{choi2021recent}%
	  \BibitemOpen
	  \bibfield  {author} {\bibinfo {author} {\bibfnamefont {K.}~\bibnamefont {{Choi}}}, \bibinfo {author} {\bibfnamefont {S.~H.}\ \bibnamefont {{Im}}},\ and\ \bibinfo {author} {\bibfnamefont {C.~S.}\ \bibnamefont {{Shin}}},\ }\bibfield  {title} {\bibinfo {title} {{Recent Progress in the Physics of Axions and Axion-Like Particles}},\ }\href {https://doi.org/10.1146/annurev-nucl-120720-031147} {\bibfield  {journal} {\bibinfo  {journal} {Annual Review of Nuclear and Particle Science}\ }\textbf {\bibinfo {volume} {71}},\ \bibinfo {pages} {225} (\bibinfo {year} {2021})}\BibitemShut {NoStop}%
	\bibitem [{\citenamefont {{Arvanitaki}}\ \emph {et~al.}(2010)\citenamefont {{Arvanitaki}}, \citenamefont {{Dimopoulos}}, \citenamefont {{Dubovsky}}, \citenamefont {{Kaloper}},\ and\ \citenamefont {{March-Russell}}}]{arvanitaki2010string}%
	  \BibitemOpen
	  \bibfield  {author} {\bibinfo {author} {\bibfnamefont {A.}~\bibnamefont {{Arvanitaki}}}, \bibinfo {author} {\bibfnamefont {S.}~\bibnamefont {{Dimopoulos}}}, \bibinfo {author} {\bibfnamefont {S.}~\bibnamefont {{Dubovsky}}}, \bibinfo {author} {\bibfnamefont {N.}~\bibnamefont {{Kaloper}}},\ and\ \bibinfo {author} {\bibfnamefont {J.}~\bibnamefont {{March-Russell}}},\ }\bibfield  {title} {\bibinfo {title} {{String axiverse}},\ }\href {https://doi.org/10.1103/PhysRevD.81.123530} {\bibfield  {journal} {\bibinfo  {journal} {\prd}\ }\textbf {\bibinfo {volume} {81}},\ \bibinfo {pages} {123530} (\bibinfo {year} {2010})}\BibitemShut {NoStop}%
	\bibitem [{\citenamefont {{Bramante}}\ and\ \citenamefont {{Raj}}(2024)}]{bramante2024dark}%
	  \BibitemOpen
	  \bibfield  {author} {\bibinfo {author} {\bibfnamefont {J.}~\bibnamefont {{Bramante}}}\ and\ \bibinfo {author} {\bibfnamefont {N.}~\bibnamefont {{Raj}}},\ }\bibfield  {title} {\bibinfo {title} {{Dark matter in compact stars}},\ }\href {https://doi.org/10.1016/j.physrep.2023.12.001} {\bibfield  {journal} {\bibinfo  {journal} {Phys.~Rep.}\ }\textbf {\bibinfo {volume} {1052}},\ \bibinfo {pages} {1} (\bibinfo {year} {2024})}\BibitemShut {NoStop}%
	\bibitem [{\citenamefont {{Leung}}\ \emph {et~al.}(2011)\citenamefont {{Leung}}, \citenamefont {{Chu}},\ and\ \citenamefont {{Lin}}}]{leung2011dark}%
	  \BibitemOpen
	  \bibfield  {author} {\bibinfo {author} {\bibfnamefont {S.~C.}\ \bibnamefont {{Leung}}}, \bibinfo {author} {\bibfnamefont {M.~C.}\ \bibnamefont {{Chu}}},\ and\ \bibinfo {author} {\bibfnamefont {L.~M.}\ \bibnamefont {{Lin}}},\ }\bibfield  {title} {\bibinfo {title} {{Dark-matter admixed neutron stars}},\ }\href {https://doi.org/10.1103/PhysRevD.84.107301} {\bibfield  {journal} {\bibinfo  {journal} {\prd}\ }\textbf {\bibinfo {volume} {84}},\ \bibinfo {eid} {107301} (\bibinfo {year} {2011})}\BibitemShut {NoStop}%
	\bibitem [{\citenamefont {{Xiang}}\ \emph {et~al.}(2014)\citenamefont {{Xiang}}, \citenamefont {{Jiang}}, \citenamefont {{Zhang}},\ and\ \citenamefont {{Yang}}}]{xiang2014effects}%
	  \BibitemOpen
	  \bibfield  {author} {\bibinfo {author} {\bibfnamefont {Q.-F.}\ \bibnamefont {{Xiang}}}, \bibinfo {author} {\bibfnamefont {W.-Z.}\ \bibnamefont {{Jiang}}}, \bibinfo {author} {\bibfnamefont {D.-R.}\ \bibnamefont {{Zhang}}},\ and\ \bibinfo {author} {\bibfnamefont {R.-Y.}\ \bibnamefont {{Yang}}},\ }\bibfield  {title} {\bibinfo {title} {{Effects of fermionic dark matter on properties of neutron stars}},\ }\href {https://doi.org/10.1103/PhysRevC.89.025803} {\bibfield  {journal} {\bibinfo  {journal} {\prc}\ }\textbf {\bibinfo {volume} {89}},\ \bibinfo {eid} {025803} (\bibinfo {year} {2014})}\BibitemShut {NoStop}%
	\bibitem [{\citenamefont {{Ellis}}\ \emph {et~al.}(2018)\citenamefont {{Ellis}}, \citenamefont {{H{\"u}tsi}}, \citenamefont {{Kannike}}, \citenamefont {{Marzola}}, \citenamefont {{Raidal}},\ and\ \citenamefont {{Vaskonen}}}]{ellis2018dark}%
	  \BibitemOpen
	  \bibfield  {author} {\bibinfo {author} {\bibfnamefont {J.}~\bibnamefont {{Ellis}}}, \bibinfo {author} {\bibfnamefont {G.}~\bibnamefont {{H{\"u}tsi}}}, \bibinfo {author} {\bibfnamefont {K.}~\bibnamefont {{Kannike}}}, \bibinfo {author} {\bibfnamefont {L.}~\bibnamefont {{Marzola}}}, \bibinfo {author} {\bibfnamefont {M.}~\bibnamefont {{Raidal}}},\ and\ \bibinfo {author} {\bibfnamefont {V.}~\bibnamefont {{Vaskonen}}},\ }\bibfield  {title} {\bibinfo {title} {{Dark matter effects on neutron star properties}},\ }\href {https://doi.org/10.1103/PhysRevD.97.123007} {\bibfield  {journal} {\bibinfo  {journal} {\prd}\ }\textbf {\bibinfo {volume} {97}},\ \bibinfo {eid} {123007} (\bibinfo {year} {2018})}\BibitemShut {NoStop}%
	\bibitem [{\citenamefont {{Leung}}\ \emph {et~al.}(2022)\citenamefont {{Leung}}, \citenamefont {{Chu}},\ and\ \citenamefont {{Lin}}}]{leung2022tidal}%
	  \BibitemOpen
	  \bibfield  {author} {\bibinfo {author} {\bibfnamefont {K.~L.}\ \bibnamefont {{Leung}}}, \bibinfo {author} {\bibfnamefont {M.~C.}\ \bibnamefont {{Chu}}},\ and\ \bibinfo {author} {\bibfnamefont {L.~M.}\ \bibnamefont {{Lin}}},\ }\bibfield  {title} {\bibinfo {title} {{Tidal deformability of dark matter admixed neutron stars}},\ }\href {https://doi.org/10.1103/PhysRevD.105.123010} {\bibfield  {journal} {\bibinfo  {journal} {\prd}\ }\textbf {\bibinfo {volume} {105}},\ \bibinfo {eid} {123010} (\bibinfo {year} {2022})}\BibitemShut {NoStop}%
	\bibitem [{\citenamefont {{Collier}}\ \emph {et~al.}(2022)\citenamefont {{Collier}}, \citenamefont {{Croon}},\ and\ \citenamefont {{Leane}}}]{collier2022tidal}%
	  \BibitemOpen
	  \bibfield  {author} {\bibinfo {author} {\bibfnamefont {M.}~\bibnamefont {{Collier}}}, \bibinfo {author} {\bibfnamefont {D.}~\bibnamefont {{Croon}}},\ and\ \bibinfo {author} {\bibfnamefont {R.~K.}\ \bibnamefont {{Leane}}},\ }\bibfield  {title} {\bibinfo {title} {{Tidal Love numbers of novel and admixed celestial objects}},\ }\href {https://doi.org/10.1103/PhysRevD.106.123027} {\bibfield  {journal} {\bibinfo  {journal} {\prd}\ }\textbf {\bibinfo {volume} {106}},\ \bibinfo {eid} {123027} (\bibinfo {year} {2022})}\BibitemShut {NoStop}%
	\bibitem [{\citenamefont {{Leung}}\ \emph {et~al.}(2012)\citenamefont {{Leung}}, \citenamefont {{Chu}},\ and\ \citenamefont {{Lin}}}]{leung2012equilibrium}%
	  \BibitemOpen
	  \bibfield  {author} {\bibinfo {author} {\bibfnamefont {S.~C.}\ \bibnamefont {{Leung}}}, \bibinfo {author} {\bibfnamefont {M.~C.}\ \bibnamefont {{Chu}}},\ and\ \bibinfo {author} {\bibfnamefont {L.~M.}\ \bibnamefont {{Lin}}},\ }\bibfield  {title} {\bibinfo {title} {{Equilibrium structure and radial oscillations of dark matter admixed neutron stars}},\ }\href {https://doi.org/10.1103/PhysRevD.85.103528} {\bibfield  {journal} {\bibinfo  {journal} {\prd}\ }\textbf {\bibinfo {volume} {85}},\ \bibinfo {eid} {103528} (\bibinfo {year} {2012})}\BibitemShut {NoStop}%
	\bibitem [{\citenamefont {{Kain}}(2021)}]{kain2021dark}%
	  \BibitemOpen
	  \bibfield  {author} {\bibinfo {author} {\bibfnamefont {B.}~\bibnamefont {{Kain}}},\ }\bibfield  {title} {\bibinfo {title} {{Dark matter admixed neutron stars}},\ }\href {https://doi.org/10.1103/PhysRevD.103.043009} {\bibfield  {journal} {\bibinfo  {journal} {\prd}\ }\textbf {\bibinfo {volume} {103}},\ \bibinfo {eid} {043009} (\bibinfo {year} {2021})}\BibitemShut {NoStop}%
	\bibitem [{\citenamefont {{Routaray}}\ \emph {et~al.}(2023)\citenamefont {{Routaray}}, \citenamefont {{Das}}, \citenamefont {{Sen}}, \citenamefont {{Kumar}}, \citenamefont {{Panotopoulos}},\ and\ \citenamefont {{Zhao}}}]{routaray2023radial}%
	  \BibitemOpen
	  \bibfield  {author} {\bibinfo {author} {\bibfnamefont {P.}~\bibnamefont {{Routaray}}}, \bibinfo {author} {\bibfnamefont {H.~C.}\ \bibnamefont {{Das}}}, \bibinfo {author} {\bibfnamefont {S.}~\bibnamefont {{Sen}}}, \bibinfo {author} {\bibfnamefont {B.}~\bibnamefont {{Kumar}}}, \bibinfo {author} {\bibfnamefont {G.}~\bibnamefont {{Panotopoulos}}},\ and\ \bibinfo {author} {\bibfnamefont {T.}~\bibnamefont {{Zhao}}},\ }\bibfield  {title} {\bibinfo {title} {{Radial oscillations of dark matter admixed neutron stars}},\ }\href {https://doi.org/10.1103/PhysRevD.107.103039} {\bibfield  {journal} {\bibinfo  {journal} {\prd}\ }\textbf {\bibinfo {volume} {107}},\ \bibinfo {eid} {103039} (\bibinfo {year} {2023})}\BibitemShut {NoStop}%
	\bibitem [{\citenamefont {{Emma}}\ \emph {et~al.}(2022)\citenamefont {{Emma}}, \citenamefont {{Schianchi}}, \citenamefont {{Pannarale}}, \citenamefont {{Sagun}},\ and\ \citenamefont {{Dietrich}}}]{emma2022numerical}%
	  \BibitemOpen
	  \bibfield  {author} {\bibinfo {author} {\bibfnamefont {M.}~\bibnamefont {{Emma}}}, \bibinfo {author} {\bibfnamefont {F.}~\bibnamefont {{Schianchi}}}, \bibinfo {author} {\bibfnamefont {F.}~\bibnamefont {{Pannarale}}}, \bibinfo {author} {\bibfnamefont {V.}~\bibnamefont {{Sagun}}},\ and\ \bibinfo {author} {\bibfnamefont {T.}~\bibnamefont {{Dietrich}}},\ }\bibfield  {title} {\bibinfo {title} {{Numerical Simulations of Dark Matter Admixed Neutron Star Binaries}},\ }\href {https://doi.org/10.3390/particles5030024} {\bibfield  {journal} {\bibinfo  {journal} {Particles}\ }\textbf {\bibinfo {volume} {5}},\ \bibinfo {pages} {273} (\bibinfo {year} {2022})}\BibitemShut {NoStop}%
	\bibitem [{\citenamefont {{Gleason}}\ \emph {et~al.}(2022)\citenamefont {{Gleason}}, \citenamefont {{Brown}},\ and\ \citenamefont {{Kain}}}]{gleason2022dynamical}%
	  \BibitemOpen
	  \bibfield  {author} {\bibinfo {author} {\bibfnamefont {T.}~\bibnamefont {{Gleason}}}, \bibinfo {author} {\bibfnamefont {B.}~\bibnamefont {{Brown}}},\ and\ \bibinfo {author} {\bibfnamefont {B.}~\bibnamefont {{Kain}}},\ }\bibfield  {title} {\bibinfo {title} {{Dynamical evolution of dark matter admixed neutron stars}},\ }\href {https://doi.org/10.1103/PhysRevD.105.023010} {\bibfield  {journal} {\bibinfo  {journal} {\prd}\ }\textbf {\bibinfo {volume} {105}},\ \bibinfo {eid} {023010} (\bibinfo {year} {2022})}\BibitemShut {NoStop}%
	\bibitem [{\citenamefont {{R{\"u}ter}}\ \emph {et~al.}(2023)\citenamefont {{R{\"u}ter}}, \citenamefont {{Sagun}}, \citenamefont {{Tichy}},\ and\ \citenamefont {{Dietrich}}}]{reuter2023quasiequilibrium}%
	  \BibitemOpen
	  \bibfield  {author} {\bibinfo {author} {\bibfnamefont {H.~R.}\ \bibnamefont {{R{\"u}ter}}}, \bibinfo {author} {\bibfnamefont {V.}~\bibnamefont {{Sagun}}}, \bibinfo {author} {\bibfnamefont {W.}~\bibnamefont {{Tichy}}},\ and\ \bibinfo {author} {\bibfnamefont {T.}~\bibnamefont {{Dietrich}}},\ }\bibfield  {title} {\bibinfo {title} {{Quasiequilibrium configurations of binary systems of dark matter admixed neutron stars}},\ }\href {https://doi.org/10.1103/PhysRevD.108.124080} {\bibfield  {journal} {\bibinfo  {journal} {\prd}\ }\textbf {\bibinfo {volume} {108}},\ \bibinfo {eid} {124080} (\bibinfo {year} {2023})}\BibitemShut {NoStop}%
	\bibitem [{\citenamefont {{Henriques}}\ \emph {et~al.}(1989)\citenamefont {{Henriques}}, \citenamefont {{Liddle}},\ and\ \citenamefont {{Moorhouse}}}]{henriques1989combined}%
	  \BibitemOpen
	  \bibfield  {author} {\bibinfo {author} {\bibfnamefont {A.~B.}\ \bibnamefont {{Henriques}}}, \bibinfo {author} {\bibfnamefont {A.~R.}\ \bibnamefont {{Liddle}}},\ and\ \bibinfo {author} {\bibfnamefont {R.~G.}\ \bibnamefont {{Moorhouse}}},\ }\bibfield  {title} {\bibinfo {title} {{Combined boson-fermion stars}},\ }\href {https://doi.org/10.1016/0370-2693(89)90623-0} {\bibfield  {journal} {\bibinfo  {journal} {Physics Letters B}\ }\textbf {\bibinfo {volume} {233}},\ \bibinfo {pages} {99} (\bibinfo {year} {1989})}\BibitemShut {NoStop}%
	\bibitem [{\citenamefont {{Henriques}}\ \emph {et~al.}(1990)\citenamefont {{Henriques}}, \citenamefont {{Liddle}},\ and\ \citenamefont {{Moorhouse}}}]{henriques1990combined}%
	  \BibitemOpen
	  \bibfield  {author} {\bibinfo {author} {\bibfnamefont {A.~B.}\ \bibnamefont {{Henriques}}}, \bibinfo {author} {\bibfnamefont {A.~R.}\ \bibnamefont {{Liddle}}},\ and\ \bibinfo {author} {\bibfnamefont {R.~G.}\ \bibnamefont {{Moorhouse}}},\ }\bibfield  {title} {\bibinfo {title} {{Combined boson-fermion stars: Configurations and stability}},\ }\href {https://doi.org/10.1016/0550-3213(90)90514-E} {\bibfield  {journal} {\bibinfo  {journal} {Nuclear Physics B}\ }\textbf {\bibinfo {volume} {337}},\ \bibinfo {pages} {737} (\bibinfo {year} {1990})}\BibitemShut {NoStop}%
	\bibitem [{\citenamefont {{Jetzer}}(1990)}]{jetzer1990stability}%
	  \BibitemOpen
	  \bibfield  {author} {\bibinfo {author} {\bibfnamefont {P.}~\bibnamefont {{Jetzer}}},\ }\bibfield  {title} {\bibinfo {title} {{Stability of combined boson-fermion stars}},\ }\href {https://doi.org/10.1016/0370-2693(90)90952-3} {\bibfield  {journal} {\bibinfo  {journal} {Physics Letters B}\ }\textbf {\bibinfo {volume} {243}},\ \bibinfo {pages} {36} (\bibinfo {year} {1990})}\BibitemShut {NoStop}%
	\bibitem [{\citenamefont {Rutherford}\ \emph {et~al.}(2023)\citenamefont {Rutherford}, \citenamefont {Raaijmakers}, \citenamefont {Prescod-Weinstein},\ and\ \citenamefont {Watts}}]{Rutherford:2023}%
	  \BibitemOpen
	  \bibfield  {author} {\bibinfo {author} {\bibfnamefont {N.}~\bibnamefont {Rutherford}}, \bibinfo {author} {\bibfnamefont {G.}~\bibnamefont {Raaijmakers}}, \bibinfo {author} {\bibfnamefont {C.}~\bibnamefont {Prescod-Weinstein}},\ and\ \bibinfo {author} {\bibfnamefont {A.}~\bibnamefont {Watts}},\ }\bibfield  {title} {\bibinfo {title} {Constraining bosonic asymmetric dark matter with neutron star mass-radius measurements},\ }\href {https://doi.org/10.1103/PhysRevD.107.103051} {\bibfield  {journal} {\bibinfo  {journal} {Phys. Rev. D}\ }\textbf {\bibinfo {volume} {107}},\ \bibinfo {pages} {103051} (\bibinfo {year} {2023})}\BibitemShut {NoStop}%
	\bibitem [{\citenamefont {{Giangrandi}}\ \emph {et~al.}(2023)\citenamefont {{Giangrandi}}, \citenamefont {{Sagun}}, \citenamefont {{Ivanytskyi}}, \citenamefont {{Provid{\^e}ncia}},\ and\ \citenamefont {{Dietrich}}}]{Giangrandi:2023}%
	  \BibitemOpen
	  \bibfield  {author} {\bibinfo {author} {\bibfnamefont {E.}~\bibnamefont {{Giangrandi}}}, \bibinfo {author} {\bibfnamefont {V.}~\bibnamefont {{Sagun}}}, \bibinfo {author} {\bibfnamefont {O.}~\bibnamefont {{Ivanytskyi}}}, \bibinfo {author} {\bibfnamefont {C.}~\bibnamefont {{Provid{\^e}ncia}}},\ and\ \bibinfo {author} {\bibfnamefont {T.}~\bibnamefont {{Dietrich}}},\ }\bibfield  {title} {\bibinfo {title} {{The Effects of Self-interacting Bosonic Dark Matter on Neutron Star Properties}},\ }\href {https://doi.org/10.3847/1538-4357/ace104} {\bibfield  {journal} {\bibinfo  {journal} {\apj}\ }\textbf {\bibinfo {volume} {953}},\ \bibinfo {eid} {115} (\bibinfo {year} {2023})},\ \Eprint {https://arxiv.org/abs/2209.10905} {arXiv:2209.10905 [astro-ph.HE]} \BibitemShut {NoStop}%
	\bibitem [{\citenamefont {{Diedrichs}}\ \emph {et~al.}(2023)\citenamefont {{Diedrichs}}, \citenamefont {{Becker}}, \citenamefont {{Jockel}}, \citenamefont {{Christian}}, \citenamefont {{Sagunski}},\ and\ \citenamefont {{Schaffner-Bielich}}}]{diedrichs2023tidal}%
	  \BibitemOpen
	  \bibfield  {author} {\bibinfo {author} {\bibfnamefont {R.~F.}\ \bibnamefont {{Diedrichs}}}, \bibinfo {author} {\bibfnamefont {N.}~\bibnamefont {{Becker}}}, \bibinfo {author} {\bibfnamefont {C.}~\bibnamefont {{Jockel}}}, \bibinfo {author} {\bibfnamefont {J.-E.}\ \bibnamefont {{Christian}}}, \bibinfo {author} {\bibfnamefont {L.}~\bibnamefont {{Sagunski}}},\ and\ \bibinfo {author} {\bibfnamefont {J.}~\bibnamefont {{Schaffner-Bielich}}},\ }\bibfield  {title} {\bibinfo {title} {{Tidal deformability of fermion-boson stars: Neutron stars admixed with ultralight dark matter}},\ }\href {https://doi.org/10.1103/PhysRevD.108.064009} {\bibfield  {journal} {\bibinfo  {journal} {\prd}\ }\textbf {\bibinfo {volume} {108}},\ \bibinfo {eid} {064009} (\bibinfo {year} {2023})}\BibitemShut {NoStop}%
	\bibitem [{\citenamefont {{Valdez-Alvarado}}\ \emph {et~al.}(2013)\citenamefont {{Valdez-Alvarado}}, \citenamefont {{Palenzuela}}, \citenamefont {{Alic}},\ and\ \citenamefont {{Ure{\~n}a-L{\'o}pez}}}]{valdezalvarado2013dynamical}%
	  \BibitemOpen
	  \bibfield  {author} {\bibinfo {author} {\bibfnamefont {S.}~\bibnamefont {{Valdez-Alvarado}}}, \bibinfo {author} {\bibfnamefont {C.}~\bibnamefont {{Palenzuela}}}, \bibinfo {author} {\bibfnamefont {D.}~\bibnamefont {{Alic}}},\ and\ \bibinfo {author} {\bibfnamefont {L.~A.}\ \bibnamefont {{Ure{\~n}a-L{\'o}pez}}},\ }\bibfield  {title} {\bibinfo {title} {{Dynamical evolution of fermion-boson stars}},\ }\href {https://doi.org/10.1103/PhysRevD.87.084040} {\bibfield  {journal} {\bibinfo  {journal} {\prd}\ }\textbf {\bibinfo {volume} {87}},\ \bibinfo {eid} {084040} (\bibinfo {year} {2013})}\BibitemShut {NoStop}%
	\bibitem [{\citenamefont {{Bezares}}\ \emph {et~al.}(2019)\citenamefont {{Bezares}}, \citenamefont {{Vigan{\`o}}},\ and\ \citenamefont {{Palenzuela}}}]{bezares2019gravitational}%
	  \BibitemOpen
	  \bibfield  {author} {\bibinfo {author} {\bibfnamefont {M.}~\bibnamefont {{Bezares}}}, \bibinfo {author} {\bibfnamefont {D.}~\bibnamefont {{Vigan{\`o}}}},\ and\ \bibinfo {author} {\bibfnamefont {C.}~\bibnamefont {{Palenzuela}}},\ }\bibfield  {title} {\bibinfo {title} {{Gravitational wave signatures of dark matter cores in binary neutron star mergers by using numerical simulations}},\ }\href {https://doi.org/10.1103/PhysRevD.100.044049} {\bibfield  {journal} {\bibinfo  {journal} {\prd}\ }\textbf {\bibinfo {volume} {100}},\ \bibinfo {pages} {044049} (\bibinfo {year} {2019})}\BibitemShut {NoStop}%
	\bibitem [{\citenamefont {{Di Giovanni}}\ \emph {et~al.}(2020)\citenamefont {{Di Giovanni}}, \citenamefont {{Fakhry}}, \citenamefont {{Sanchis-Gual}}, \citenamefont {{Degollado}},\ and\ \citenamefont {{Font}}}]{digiovanni2020dynamical}%
	  \BibitemOpen
	  \bibfield  {author} {\bibinfo {author} {\bibfnamefont {F.}~\bibnamefont {{Di Giovanni}}}, \bibinfo {author} {\bibfnamefont {S.}~\bibnamefont {{Fakhry}}}, \bibinfo {author} {\bibfnamefont {N.}~\bibnamefont {{Sanchis-Gual}}}, \bibinfo {author} {\bibfnamefont {J.~C.}\ \bibnamefont {{Degollado}}},\ and\ \bibinfo {author} {\bibfnamefont {J.~A.}\ \bibnamefont {{Font}}},\ }\bibfield  {title} {\bibinfo {title} {{Dynamical formation and stability of fermion-boson stars}},\ }\href {https://doi.org/10.1103/PhysRevD.102.084063} {\bibfield  {journal} {\bibinfo  {journal} {\prd}\ }\textbf {\bibinfo {volume} {102}},\ \bibinfo {pages} {084063} (\bibinfo {year} {2020})}\BibitemShut {NoStop}%
	\bibitem [{\citenamefont {{Lee}}\ \emph {et~al.}(2021)\citenamefont {{Lee}}, \citenamefont {{Chu}},\ and\ \citenamefont {{Lin}}}]{lee2021could}%
	  \BibitemOpen
	  \bibfield  {author} {\bibinfo {author} {\bibfnamefont {B.~K.~K.}\ \bibnamefont {{Lee}}}, \bibinfo {author} {\bibfnamefont {M.-C.}\ \bibnamefont {{Chu}}},\ and\ \bibinfo {author} {\bibfnamefont {L.-M.}\ \bibnamefont {{Lin}}},\ }\bibfield  {title} {\bibinfo {title} {{Could the GW190814 Secondary Component Be a Bosonic Dark Matter Admixed Compact Star?}},\ }\href {https://doi.org/10.3847/1538-4357/ac2735} {\bibfield  {journal} {\bibinfo  {journal} {\apj}\ }\textbf {\bibinfo {volume} {922}},\ \bibinfo {eid} {242} (\bibinfo {year} {2021})}\BibitemShut {NoStop}%
	\bibitem [{\citenamefont {{Di Giovanni}}\ \emph {et~al.}(2022)\citenamefont {{Di Giovanni}}, \citenamefont {{Sanchis-Gual}}, \citenamefont {{Cerd{\'a}-Dur{\'a}n}},\ and\ \citenamefont {{Font}}}]{digiovanni2022can}%
	  \BibitemOpen
	  \bibfield  {author} {\bibinfo {author} {\bibfnamefont {F.}~\bibnamefont {{Di Giovanni}}}, \bibinfo {author} {\bibfnamefont {N.}~\bibnamefont {{Sanchis-Gual}}}, \bibinfo {author} {\bibfnamefont {P.}~\bibnamefont {{Cerd{\'a}-Dur{\'a}n}}},\ and\ \bibinfo {author} {\bibfnamefont {J.~A.}\ \bibnamefont {{Font}}},\ }\bibfield  {title} {\bibinfo {title} {{Can fermion-boson stars reconcile multimessenger observations of compact stars?}},\ }\href {https://doi.org/10.1103/PhysRevD.105.063005} {\bibfield  {journal} {\bibinfo  {journal} {\prd}\ }\textbf {\bibinfo {volume} {105}},\ \bibinfo {eid} {063005} (\bibinfo {year} {2022})}\BibitemShut {NoStop}%
	\bibitem [{\citenamefont {Nyhan}\ and\ \citenamefont {Kain}(2022)}]{Nyhan:2022}%
	  \BibitemOpen
	  \bibfield  {author} {\bibinfo {author} {\bibfnamefont {J.~E.}\ \bibnamefont {Nyhan}}\ and\ \bibinfo {author} {\bibfnamefont {B.}~\bibnamefont {Kain}},\ }\bibfield  {title} {\bibinfo {title} {Dynamical evolution of fermion-boson stars with realistic equations of state},\ }\href {https://doi.org/10.1103/PhysRevD.105.123016} {\bibfield  {journal} {\bibinfo  {journal} {Phys. Rev. D}\ }\textbf {\bibinfo {volume} {105}},\ \bibinfo {pages} {123016} (\bibinfo {year} {2022})}\BibitemShut {NoStop}%
	\bibitem [{\citenamefont {Prabhu}(2021)}]{Prabhu:2021}%
	  \BibitemOpen
	  \bibfield  {author} {\bibinfo {author} {\bibfnamefont {A.}~\bibnamefont {Prabhu}},\ }\bibfield  {title} {\bibinfo {title} {Axion production in pulsar magnetosphere gaps},\ }\href {https://doi.org/10.1103/PhysRevD.104.055038} {\bibfield  {journal} {\bibinfo  {journal} {Phys. Rev. D}\ }\textbf {\bibinfo {volume} {104}},\ \bibinfo {pages} {055038} (\bibinfo {year} {2021})}\BibitemShut {NoStop}%
	\bibitem [{\citenamefont {Noordhuis}\ \emph {et~al.}(2023)\citenamefont {Noordhuis}, \citenamefont {Prabhu}, \citenamefont {Witte}, \citenamefont {Chen}, \citenamefont {Cruz},\ and\ \citenamefont {Weniger}}]{Noordhuis:2023}%
	  \BibitemOpen
	  \bibfield  {author} {\bibinfo {author} {\bibfnamefont {D.}~\bibnamefont {Noordhuis}}, \bibinfo {author} {\bibfnamefont {A.}~\bibnamefont {Prabhu}}, \bibinfo {author} {\bibfnamefont {S.~J.}\ \bibnamefont {Witte}}, \bibinfo {author} {\bibfnamefont {A.~Y.}\ \bibnamefont {Chen}}, \bibinfo {author} {\bibfnamefont {F.}~\bibnamefont {Cruz}},\ and\ \bibinfo {author} {\bibfnamefont {C.}~\bibnamefont {Weniger}},\ }\bibfield  {title} {\bibinfo {title} {Novel constraints on axions produced in pulsar polar-cap cascades},\ }\href {https://doi.org/10.1103/PhysRevLett.131.111004} {\bibfield  {journal} {\bibinfo  {journal} {Phys. Rev. Lett.}\ }\textbf {\bibinfo {volume} {131}},\ \bibinfo {pages} {111004} (\bibinfo {year} {2023})}\BibitemShut {NoStop}%
	\bibitem [{\citenamefont {{Yagi}}\ and\ \citenamefont {{Yunes}}(2017)}]{yagi2017approximate}%
	  \BibitemOpen
	  \bibfield  {author} {\bibinfo {author} {\bibfnamefont {K.}~\bibnamefont {{Yagi}}}\ and\ \bibinfo {author} {\bibfnamefont {N.}~\bibnamefont {{Yunes}}},\ }\bibfield  {title} {\bibinfo {title} {{Approximate universal relations for neutron stars and quark stars}},\ }\href {https://doi.org/10.1016/j.physrep.2017.03.002} {\bibfield  {journal} {\bibinfo  {journal} {Phys.~Rep.}\ }\textbf {\bibinfo {volume} {681}},\ \bibinfo {pages} {1} (\bibinfo {year} {2017})}\BibitemShut {NoStop}%
	\bibitem [{\citenamefont {Doneva}\ and\ \citenamefont {Pappas}(2018)}]{Doneva:2018}%
	  \BibitemOpen
	  \bibfield  {author} {\bibinfo {author} {\bibfnamefont {D.~D.}\ \bibnamefont {Doneva}}\ and\ \bibinfo {author} {\bibfnamefont {G.}~\bibnamefont {Pappas}},\ }\bibinfo {title} {Universal relations and alternative gravity theories},\ in\ \href {https://doi.org/10.1007/978-3-319-97616-7_13} {\emph {\bibinfo {booktitle} {The Physics and Astrophysics of Neutron Stars}}},\ \bibinfo {editor} {edited by\ \bibinfo {editor} {\bibfnamefont {L.}~\bibnamefont {Rezzolla}}, \bibinfo {editor} {\bibfnamefont {P.}~\bibnamefont {Pizzochero}}, \bibinfo {editor} {\bibfnamefont {D.~I.}\ \bibnamefont {Jones}}, \bibinfo {editor} {\bibfnamefont {N.}~\bibnamefont {Rea}},\ and\ \bibinfo {editor} {\bibfnamefont {I.}~\bibnamefont {Vida{\~{n}}a}}}\ (\bibinfo  {publisher} {Springer International Publishing},\ \bibinfo {address} {Cham},\ \bibinfo {year} {2018})\ pp.\ \bibinfo {pages} {737--806}\BibitemShut {NoStop}%
	\bibitem [{\citenamefont {Lau}\ \emph {et~al.}(2010)\citenamefont {Lau}, \citenamefont {Leung},\ and\ \citenamefont {Lin}}]{Lau:2010}%
	  \BibitemOpen
	  \bibfield  {author} {\bibinfo {author} {\bibfnamefont {H.~K.}\ \bibnamefont {Lau}}, \bibinfo {author} {\bibfnamefont {P.~T.}\ \bibnamefont {Leung}},\ and\ \bibinfo {author} {\bibfnamefont {L.~M.}\ \bibnamefont {Lin}},\ }\bibfield  {title} {\bibinfo {title} {Inferring physical parameters of compact stars from their f-mode gravitational wave signals},\ }\href {https://doi.org/10.1088/0004-637x/714/2/1234} {\bibfield  {journal} {\bibinfo  {journal} {Astrophys. J.}\ }\textbf {\bibinfo {volume} {714}},\ \bibinfo {pages} {1234} (\bibinfo {year} {2010})}\BibitemShut {NoStop}%
	\bibitem [{\citenamefont {{Yagi}}\ and\ \citenamefont {{Yunes}}(2013)}]{yagi2013iloveq}%
	  \BibitemOpen
	  \bibfield  {author} {\bibinfo {author} {\bibfnamefont {K.}~\bibnamefont {{Yagi}}}\ and\ \bibinfo {author} {\bibfnamefont {N.}~\bibnamefont {{Yunes}}},\ }\bibfield  {title} {\bibinfo {title} {{I-Love-Q relations in neutron stars and their applications to astrophysics, gravitational waves, and fundamental physics}},\ }\href {https://doi.org/10.1103/PhysRevD.88.023009} {\bibfield  {journal} {\bibinfo  {journal} {\prd}\ }\textbf {\bibinfo {volume} {88}},\ \bibinfo {eid} {023009} (\bibinfo {year} {2013})}\BibitemShut {NoStop}%
	\bibitem [{\citenamefont {{Kumar}}\ and\ \citenamefont {{Landry}}(2019)}]{kumar2019inferring}%
	  \BibitemOpen
	  \bibfield  {author} {\bibinfo {author} {\bibfnamefont {B.}~\bibnamefont {{Kumar}}}\ and\ \bibinfo {author} {\bibfnamefont {P.}~\bibnamefont {{Landry}}},\ }\bibfield  {title} {\bibinfo {title} {{Inferring neutron star properties from GW170817 with universal relations}},\ }\href {https://doi.org/10.1103/PhysRevD.99.123026} {\bibfield  {journal} {\bibinfo  {journal} {\prd}\ }\textbf {\bibinfo {volume} {99}},\ \bibinfo {pages} {123026} (\bibinfo {year} {2019})}\BibitemShut {NoStop}%
	\bibitem [{\citenamefont {Chan}\ \emph {et~al.}(2014)\citenamefont {Chan}, \citenamefont {Sham}, \citenamefont {Leung},\ and\ \citenamefont {Lin}}]{Chan:2014}%
	  \BibitemOpen
	  \bibfield  {author} {\bibinfo {author} {\bibfnamefont {T.~K.}\ \bibnamefont {Chan}}, \bibinfo {author} {\bibfnamefont {Y.-H.}\ \bibnamefont {Sham}}, \bibinfo {author} {\bibfnamefont {P.~T.}\ \bibnamefont {Leung}},\ and\ \bibinfo {author} {\bibfnamefont {L.-M.}\ \bibnamefont {Lin}},\ }\bibfield  {title} {\bibinfo {title} {Multipolar universal relations between $f$-mode frequency and tidal deformability of compact stars},\ }\href {https://doi.org/10.1103/PhysRevD.90.124023} {\bibfield  {journal} {\bibinfo  {journal} {Phys. Rev. D}\ }\textbf {\bibinfo {volume} {90}},\ \bibinfo {pages} {124023} (\bibinfo {year} {2014})}\BibitemShut {NoStop}%
	\bibitem [{\citenamefont {{Cronin}}\ \emph {et~al.}(2023)\citenamefont {{Cronin}}, \citenamefont {{Zhang}},\ and\ \citenamefont {{Kain}}}]{cronin2023rotating}%
	  \BibitemOpen
	  \bibfield  {author} {\bibinfo {author} {\bibfnamefont {J.}~\bibnamefont {{Cronin}}}, \bibinfo {author} {\bibfnamefont {X.}~\bibnamefont {{Zhang}}},\ and\ \bibinfo {author} {\bibfnamefont {B.}~\bibnamefont {{Kain}}},\ }\bibfield  {title} {\bibinfo {title} {{Rotating dark matter admixed neutron stars}},\ }\href {https://doi.org/10.1103/PhysRevD.108.103016} {\bibfield  {journal} {\bibinfo  {journal} {\prd}\ }\textbf {\bibinfo {volume} {108}},\ \bibinfo {pages} {103016} (\bibinfo {year} {2023})}\BibitemShut {NoStop}%
	\bibitem [{\citenamefont {{Hartle}}(1967)}]{hartle1967slowly}%
	  \BibitemOpen
	  \bibfield  {author} {\bibinfo {author} {\bibfnamefont {J.~B.}\ \bibnamefont {{Hartle}}},\ }\bibfield  {title} {\bibinfo {title} {{Slowly Rotating Relativistic Stars. I. Equations of Structure}},\ }\href {https://doi.org/10.1086/149400} {\bibfield  {journal} {\bibinfo  {journal} {\apj}\ }\textbf {\bibinfo {volume} {150}},\ \bibinfo {pages} {1005} (\bibinfo {year} {1967})}\BibitemShut {NoStop}%
	\bibitem [{\citenamefont {{Akmal}}\ \emph {et~al.}(1998)\citenamefont {{Akmal}}, \citenamefont {{Pandharipande}},\ and\ \citenamefont {{Ravenhall}}}]{akmal1998equation}%
	  \BibitemOpen
	  \bibfield  {author} {\bibinfo {author} {\bibfnamefont {A.}~\bibnamefont {{Akmal}}}, \bibinfo {author} {\bibfnamefont {V.~R.}\ \bibnamefont {{Pandharipande}}},\ and\ \bibinfo {author} {\bibfnamefont {D.~G.}\ \bibnamefont {{Ravenhall}}},\ }\bibfield  {title} {\bibinfo {title} {{Equation of state of nucleon matter and neutron star structure}},\ }\href {https://doi.org/10.1103/PhysRevC.58.1804} {\bibfield  {journal} {\bibinfo  {journal} {\prc}\ }\textbf {\bibinfo {volume} {58}},\ \bibinfo {pages} {1804} (\bibinfo {year} {1998})}\BibitemShut {NoStop}%
	\bibitem [{\citenamefont {{Typel}}\ \emph {et~al.}(2010)\citenamefont {{Typel}}, \citenamefont {{R{\"o}pke}}, \citenamefont {{Kl{\"a}hn}}, \citenamefont {{Blaschke}},\ and\ \citenamefont {{Wolter}}}]{typel2010composition}%
	  \BibitemOpen
	  \bibfield  {author} {\bibinfo {author} {\bibfnamefont {S.}~\bibnamefont {{Typel}}}, \bibinfo {author} {\bibfnamefont {G.}~\bibnamefont {{R{\"o}pke}}}, \bibinfo {author} {\bibfnamefont {T.}~\bibnamefont {{Kl{\"a}hn}}}, \bibinfo {author} {\bibfnamefont {D.}~\bibnamefont {{Blaschke}}},\ and\ \bibinfo {author} {\bibfnamefont {H.~H.}\ \bibnamefont {{Wolter}}},\ }\bibfield  {title} {\bibinfo {title} {{Composition and thermodynamics of nuclear matter with light clusters}},\ }\href {https://doi.org/10.1103/PhysRevC.81.015803} {\bibfield  {journal} {\bibinfo  {journal} {\prc}\ }\textbf {\bibinfo {volume} {81}},\ \bibinfo {eid} {015803} (\bibinfo {year} {2010})}\BibitemShut {NoStop}%
	\bibitem [{\citenamefont {{Shapiro}}\ and\ \citenamefont {{Teukolsky}}(1983)}]{shapiro1983black}%
	  \BibitemOpen
	  \bibfield  {author} {\bibinfo {author} {\bibfnamefont {S.~L.}\ \bibnamefont {{Shapiro}}}\ and\ \bibinfo {author} {\bibfnamefont {S.~A.}\ \bibnamefont {{Teukolsky}}},\ }\href {https://doi.org/10.1002/9783527617661} {\emph {\bibinfo {title} {{Black holes, white dwarfs and neutron stars. The physics of compact objects}}}}\ (\bibinfo  {publisher} {John Wiley \& Sons},\ \bibinfo {year} {1983})\BibitemShut {NoStop}%
	\bibitem [{\citenamefont {{Lee}}\ and\ \citenamefont {{Pang}}(1989)}]{lee1989stability}%
	  \BibitemOpen
	  \bibfield  {author} {\bibinfo {author} {\bibfnamefont {T.~D.}\ \bibnamefont {{Lee}}}\ and\ \bibinfo {author} {\bibfnamefont {Y.}~\bibnamefont {{Pang}}},\ }\bibfield  {title} {\bibinfo {title} {Stability of mini-boson stars},\ }\href {https://doi.org/10.1016/0550-3213(89)90365-9} {\bibfield  {journal} {\bibinfo  {journal} {Nuclear Physics B}\ }\textbf {\bibinfo {volume} {315}},\ \bibinfo {pages} {477} (\bibinfo {year} {1989})}\BibitemShut {NoStop}%
	\bibitem [{\citenamefont {{Liebling}}\ and\ \citenamefont {{Palenzuela}}(2023)}]{liebling2023dynamical}%
	  \BibitemOpen
	  \bibfield  {author} {\bibinfo {author} {\bibfnamefont {S.~L.}\ \bibnamefont {{Liebling}}}\ and\ \bibinfo {author} {\bibfnamefont {C.}~\bibnamefont {{Palenzuela}}},\ }\bibfield  {title} {\bibinfo {title} {{Dynamical boson stars}},\ }\href {https://doi.org/10.1007/s41114-023-00043-4} {\bibfield  {journal} {\bibinfo  {journal} {Living Reviews in Relativity}\ }\textbf {\bibinfo {volume} {26}},\ \bibinfo {eid} {1} (\bibinfo {year} {2023})}\BibitemShut {NoStop}%
	\bibitem [{\citenamefont {{Schunck}}\ and\ \citenamefont {{Mielke}}(1998)}]{schunck1998rotating}%
	  \BibitemOpen
	  \bibfield  {author} {\bibinfo {author} {\bibfnamefont {F.~E.}\ \bibnamefont {{Schunck}}}\ and\ \bibinfo {author} {\bibfnamefont {E.~W.}\ \bibnamefont {{Mielke}}},\ }\bibfield  {title} {\bibinfo {title} {{Rotating boson star as an effective mass torus in general relativity}},\ }\href {https://doi.org/10.1016/S0375-9601(98)00778-6} {\bibfield  {journal} {\bibinfo  {journal} {Physics Letters A}\ }\textbf {\bibinfo {volume} {249}},\ \bibinfo {pages} {389} (\bibinfo {year} {1998})}\BibitemShut {NoStop}%
	\bibitem [{\citenamefont {{Kobayashi}}\ \emph {et~al.}(1994)\citenamefont {{Kobayashi}}, \citenamefont {{Kasai}},\ and\ \citenamefont {{Futamase}}}]{kobayashi1994does}%
	  \BibitemOpen
	  \bibfield  {author} {\bibinfo {author} {\bibfnamefont {Y.-S.}\ \bibnamefont {{Kobayashi}}}, \bibinfo {author} {\bibfnamefont {M.}~\bibnamefont {{Kasai}}},\ and\ \bibinfo {author} {\bibfnamefont {T.}~\bibnamefont {{Futamase}}},\ }\bibfield  {title} {\bibinfo {title} {{Does a boson star rotate\textbackslash?}},\ }\href {https://doi.org/10.1103/PhysRevD.50.7721} {\bibfield  {journal} {\bibinfo  {journal} {\prd}\ }\textbf {\bibinfo {volume} {50}},\ \bibinfo {pages} {7721} (\bibinfo {year} {1994})}\BibitemShut {NoStop}%
	\bibitem [{\citenamefont {{de Sousa}}\ \emph {et~al.}(2001)\citenamefont {{de Sousa}}, \citenamefont {{Silveira}},\ and\ \citenamefont {{Fang}}}]{deSousa2001slowly}%
	  \BibitemOpen
	  \bibfield  {author} {\bibinfo {author} {\bibfnamefont {C.~M.~G.}\ \bibnamefont {{de Sousa}}}, \bibinfo {author} {\bibfnamefont {V.}~\bibnamefont {{Silveira}}},\ and\ \bibinfo {author} {\bibfnamefont {L.~Z.}\ \bibnamefont {{Fang}}},\ }\bibfield  {title} {\bibinfo {title} {{Slowly Rotating Boson-Fermion Star}},\ }\href {https://doi.org/10.1142/S0218271801001360} {\bibfield  {journal} {\bibinfo  {journal} {International Journal of Modern Physics D}\ }\textbf {\bibinfo {volume} {10}},\ \bibinfo {pages} {881} (\bibinfo {year} {2001})}\BibitemShut {NoStop}%
	\bibitem [{\citenamefont {{Hinderer}}(2008)}]{hinderer2008tidal}%
	  \BibitemOpen
	  \bibfield  {author} {\bibinfo {author} {\bibfnamefont {T.}~\bibnamefont {{Hinderer}}},\ }\bibfield  {title} {\bibinfo {title} {{Tidal Love Numbers of Neutron Stars}},\ }\href {https://doi.org/10.1086/533487} {\bibfield  {journal} {\bibinfo  {journal} {\apj}\ }\textbf {\bibinfo {volume} {677}},\ \bibinfo {pages} {1216} (\bibinfo {year} {2008})}\BibitemShut {NoStop}%
	\bibitem [{\citenamefont {{Sennett}}\ \emph {et~al.}(2017)\citenamefont {{Sennett}}, \citenamefont {{Hinderer}}, \citenamefont {{Steinhoff}}, \citenamefont {{Buonanno}},\ and\ \citenamefont {{Ossokine}}}]{sennett2017distinguishing}%
	  \BibitemOpen
	  \bibfield  {author} {\bibinfo {author} {\bibfnamefont {N.}~\bibnamefont {{Sennett}}}, \bibinfo {author} {\bibfnamefont {T.}~\bibnamefont {{Hinderer}}}, \bibinfo {author} {\bibfnamefont {J.}~\bibnamefont {{Steinhoff}}}, \bibinfo {author} {\bibfnamefont {A.}~\bibnamefont {{Buonanno}}},\ and\ \bibinfo {author} {\bibfnamefont {S.}~\bibnamefont {{Ossokine}}},\ }\bibfield  {title} {\bibinfo {title} {{Distinguishing boson stars from black holes and neutron stars from tidal interactions in inspiraling binary systems}},\ }\href {https://doi.org/10.1103/PhysRevD.96.024002} {\bibfield  {journal} {\bibinfo  {journal} {\prd}\ }\textbf {\bibinfo {volume} {96}},\ \bibinfo {eid} {024002} (\bibinfo {year} {2017})}\BibitemShut {NoStop}%
	\bibitem [{\citenamefont {{Kiziltan}}\ \emph {et~al.}(2013)\citenamefont {{Kiziltan}}, \citenamefont {{Kottas}}, \citenamefont {{De Yoreo}},\ and\ \citenamefont {{Thorsett}}}]{kiziltan2013the}%
	  \BibitemOpen
	  \bibfield  {author} {\bibinfo {author} {\bibfnamefont {B.}~\bibnamefont {{Kiziltan}}}, \bibinfo {author} {\bibfnamefont {A.}~\bibnamefont {{Kottas}}}, \bibinfo {author} {\bibfnamefont {M.}~\bibnamefont {{De Yoreo}}},\ and\ \bibinfo {author} {\bibfnamefont {S.~E.}\ \bibnamefont {{Thorsett}}},\ }\bibfield  {title} {\bibinfo {title} {{The Neutron Star Mass Distribution}},\ }\href {https://doi.org/10.1088/0004-637X/778/1/66} {\bibfield  {journal} {\bibinfo  {journal} {\apj}\ }\textbf {\bibinfo {volume} {778}},\ \bibinfo {pages} {66} (\bibinfo {year} {2013})}\BibitemShut {NoStop}%
	\bibitem [{\citenamefont {{Landry}}\ and\ \citenamefont {{Read}}(2021)}]{landry2021the}%
	  \BibitemOpen
	  \bibfield  {author} {\bibinfo {author} {\bibfnamefont {P.}~\bibnamefont {{Landry}}}\ and\ \bibinfo {author} {\bibfnamefont {J.~S.}\ \bibnamefont {{Read}}},\ }\bibfield  {title} {\bibinfo {title} {{The Mass Distribution of Neutron Stars in Gravitational-wave Binaries}},\ }\href {https://doi.org/10.3847/2041-8213/ac2f3e} {\bibfield  {journal} {\bibinfo  {journal} {ApJ}\ }\textbf {\bibinfo {volume} {921}},\ \bibinfo {pages} {L25} (\bibinfo {year} {2021})}\BibitemShut {NoStop}%
	\bibitem [{\citenamefont {{Abbott}}\ \emph {et~al.}(2019)\citenamefont {{Abbott}}, \citenamefont {{Abbott}}, \citenamefont {{Abbott}}, \citenamefont {{Acernese}}, \citenamefont {{Ackley}}, \citenamefont {{Adams}}, \citenamefont {{Adams}}, \citenamefont {{Addesso}}, \citenamefont {{Adhikari}}, \citenamefont {{Adya}} \emph {et~al.}}]{abbott2019properties}%
	  \BibitemOpen
	  \bibfield  {author} {\bibinfo {author} {\bibfnamefont {B.~P.}\ \bibnamefont {{Abbott}}}, \bibinfo {author} {\bibfnamefont {R.}~\bibnamefont {{Abbott}}}, \bibinfo {author} {\bibfnamefont {T.~D.}\ \bibnamefont {{Abbott}}}, \bibinfo {author} {\bibfnamefont {F.}~\bibnamefont {{Acernese}}}, \bibinfo {author} {\bibfnamefont {K.}~\bibnamefont {{Ackley}}}, \bibinfo {author} {\bibfnamefont {C.}~\bibnamefont {{Adams}}}, \bibinfo {author} {\bibfnamefont {T.}~\bibnamefont {{Adams}}}, \bibinfo {author} {\bibfnamefont {P.}~\bibnamefont {{Addesso}}}, \bibinfo {author} {\bibfnamefont {R.~X.}\ \bibnamefont {{Adhikari}}}, \bibinfo {author} {\bibfnamefont {V.~B.}\ \bibnamefont {{Adya}}}, \emph {et~al.},\ }\bibfield  {title} {\bibinfo {title} {{Properties of the Binary Neutron Star Merger GW170817}},\ }\href {https://doi.org/10.1103/PhysRevX.9.011001} {\bibfield  {journal} {\bibinfo  {journal} {Phys. Rev. X}\ }\textbf {\bibinfo {volume} {9}},\ \bibinfo {eid} {011001} (\bibinfo {year} {2019})}\BibitemShut {NoStop}%
	\bibitem [{\citenamefont {{Hempel}}\ and\ \citenamefont {{Schaffner-Bielich}}(2010)}]{hempel2010a}%
	  \BibitemOpen
	  \bibfield  {author} {\bibinfo {author} {\bibfnamefont {M.}~\bibnamefont {{Hempel}}}\ and\ \bibinfo {author} {\bibfnamefont {J.}~\bibnamefont {{Schaffner-Bielich}}},\ }\bibfield  {title} {\bibinfo {title} {{A statistical model for a complete supernova equation of state}},\ }\href {https://doi.org/10.1016/j.nuclphysa.2010.02.010} {\bibfield  {journal} {\bibinfo  {journal} {Nucl. Phys. A}\ }\textbf {\bibinfo {volume} {837}},\ \bibinfo {pages} {210} (\bibinfo {year} {2010})}\BibitemShut {NoStop}%
	\bibitem [{\citenamefont {Lackey}\ \emph {et~al.}(2019)\citenamefont {Lackey}, \citenamefont {P\"urrer}, \citenamefont {Taracchini},\ and\ \citenamefont {Marsat}}]{Lackey:2019}%
	  \BibitemOpen
	  \bibfield  {author} {\bibinfo {author} {\bibfnamefont {B.~D.}\ \bibnamefont {Lackey}}, \bibinfo {author} {\bibfnamefont {M.}~\bibnamefont {P\"urrer}}, \bibinfo {author} {\bibfnamefont {A.}~\bibnamefont {Taracchini}},\ and\ \bibinfo {author} {\bibfnamefont {S.}~\bibnamefont {Marsat}},\ }\bibfield  {title} {\bibinfo {title} {Surrogate model for an aligned-spin effective-one-body waveform model of binary neutron star inspirals using gaussian process regression},\ }\href {https://doi.org/10.1103/PhysRevD.100.024002} {\bibfield  {journal} {\bibinfo  {journal} {Phys. Rev. D}\ }\textbf {\bibinfo {volume} {100}},\ \bibinfo {pages} {024002} (\bibinfo {year} {2019})}\BibitemShut {NoStop}%
	\bibitem [{\citenamefont {Schmidt}\ and\ \citenamefont {Hinderer}(2019)}]{Schmidt:2019}%
	  \BibitemOpen
	  \bibfield  {author} {\bibinfo {author} {\bibfnamefont {P.}~\bibnamefont {Schmidt}}\ and\ \bibinfo {author} {\bibfnamefont {T.}~\bibnamefont {Hinderer}},\ }\bibfield  {title} {\bibinfo {title} {Frequency domain model of $f$-mode dynamic tides in gravitational waveforms from compact binary inspirals},\ }\href {https://doi.org/10.1103/PhysRevD.100.021501} {\bibfield  {journal} {\bibinfo  {journal} {Phys. Rev. D}\ }\textbf {\bibinfo {volume} {100}},\ \bibinfo {pages} {021501} (\bibinfo {year} {2019})}\BibitemShut {NoStop}%
	\bibitem [{\citenamefont {Barkett}\ \emph {et~al.}(2020)\citenamefont {Barkett}, \citenamefont {Chen}, \citenamefont {Scheel},\ and\ \citenamefont {Varma}}]{Barkett:2020}%
	  \BibitemOpen
	  \bibfield  {author} {\bibinfo {author} {\bibfnamefont {K.}~\bibnamefont {Barkett}}, \bibinfo {author} {\bibfnamefont {Y.}~\bibnamefont {Chen}}, \bibinfo {author} {\bibfnamefont {M.~A.}\ \bibnamefont {Scheel}},\ and\ \bibinfo {author} {\bibfnamefont {V.}~\bibnamefont {Varma}},\ }\bibfield  {title} {\bibinfo {title} {Gravitational waveforms of binary neutron star inspirals using post-newtonian tidal splicing},\ }\href {https://doi.org/10.1103/PhysRevD.102.024031} {\bibfield  {journal} {\bibinfo  {journal} {Phys. Rev. D}\ }\textbf {\bibinfo {volume} {102}},\ \bibinfo {pages} {024031} (\bibinfo {year} {2020})}\BibitemShut {NoStop}%
	\bibitem [{\citenamefont {{Andersson}}\ and\ \citenamefont {{Pnigouras}}(2021)}]{Andersson:2021}%
	  \BibitemOpen
	  \bibfield  {author} {\bibinfo {author} {\bibfnamefont {N.}~\bibnamefont {{Andersson}}}\ and\ \bibinfo {author} {\bibfnamefont {P.}~\bibnamefont {{Pnigouras}}},\ }\bibfield  {title} {\bibinfo {title} {{The phenomenology of dynamical neutron star tides}},\ }\href {https://doi.org/10.1093/mnras/stab371} {\bibfield  {journal} {\bibinfo  {journal} {MNRAS}\ }\textbf {\bibinfo {volume} {503}},\ \bibinfo {pages} {533} (\bibinfo {year} {2021})},\ \Eprint {https://arxiv.org/abs/1905.00012} {arXiv:1905.00012 [gr-qc]} \BibitemShut {NoStop}%
	\bibitem [{\citenamefont {{Sham}}\ \emph {et~al.}(2014)\citenamefont {{Sham}}, \citenamefont {{Lin}},\ and\ \citenamefont {{Leung}}}]{Sham:2014}%
	  \BibitemOpen
	  \bibfield  {author} {\bibinfo {author} {\bibfnamefont {Y.~H.}\ \bibnamefont {{Sham}}}, \bibinfo {author} {\bibfnamefont {L.~M.}\ \bibnamefont {{Lin}}},\ and\ \bibinfo {author} {\bibfnamefont {P.~T.}\ \bibnamefont {{Leung}}},\ }\bibfield  {title} {\bibinfo {title} {{Testing Universal Relations of Neutron Stars with a Nonlinear Matter-Gravity Coupling Theory}},\ }\href {https://doi.org/10.1088/0004-637X/781/2/66} {\bibfield  {journal} {\bibinfo  {journal} {\apj}\ }\textbf {\bibinfo {volume} {781}},\ \bibinfo {eid} {66} (\bibinfo {year} {2014})},\ \Eprint {https://arxiv.org/abs/1312.1011} {arXiv:1312.1011 [gr-qc]} \BibitemShut {NoStop}%
	\bibitem [{\citenamefont {{Berti}}\ \emph {et~al.}(2015)\citenamefont {{Berti}}, \citenamefont {{Barausse}}, \citenamefont {{Cardoso}}, \citenamefont {{Gualtieri}}, \citenamefont {{Pani}}, \citenamefont {{Sperhake}}, \citenamefont {{Stein}}, \citenamefont {{Wex}}, \citenamefont {{Yagi}}, \citenamefont {{Baker}} \emph {et~al.}}]{berti2015testing}%
	  \BibitemOpen
	  \bibfield  {author} {\bibinfo {author} {\bibfnamefont {E.}~\bibnamefont {{Berti}}}, \bibinfo {author} {\bibfnamefont {E.}~\bibnamefont {{Barausse}}}, \bibinfo {author} {\bibfnamefont {V.}~\bibnamefont {{Cardoso}}}, \bibinfo {author} {\bibfnamefont {L.}~\bibnamefont {{Gualtieri}}}, \bibinfo {author} {\bibfnamefont {P.}~\bibnamefont {{Pani}}}, \bibinfo {author} {\bibfnamefont {U.}~\bibnamefont {{Sperhake}}}, \bibinfo {author} {\bibfnamefont {L.~C.}\ \bibnamefont {{Stein}}}, \bibinfo {author} {\bibfnamefont {N.}~\bibnamefont {{Wex}}}, \bibinfo {author} {\bibfnamefont {K.}~\bibnamefont {{Yagi}}}, \bibinfo {author} {\bibfnamefont {T.}~\bibnamefont {{Baker}}}, \emph {et~al.},\ }\bibfield  {title} {\bibinfo {title} {{Testing general relativity with present and future astrophysical observations}},\ }\href {https://doi.org/10.1088/0264-9381/32/24/243001} {\bibfield  {journal} {\bibinfo  {journal} {Classical and Quantum Gravity}\ }\textbf {\bibinfo {volume} {32}},\ \bibinfo {pages} {243001} (\bibinfo {year} {2015})}\BibitemShut {NoStop}%
	\bibitem [{\citenamefont {{Pani}}(2015)}]{pani2015iloveq}%
	  \BibitemOpen
	  \bibfield  {author} {\bibinfo {author} {\bibfnamefont {P.}~\bibnamefont {{Pani}}},\ }\bibfield  {title} {\bibinfo {title} {{I-Love-Q relations for gravastars and the approach to the black-hole limit}},\ }\href {https://doi.org/10.1103/PhysRevD.92.124030} {\bibfield  {journal} {\bibinfo  {journal} {\prd}\ }\textbf {\bibinfo {volume} {92}},\ \bibinfo {eid} {124030} (\bibinfo {year} {2015})}\BibitemShut {NoStop}%
	\bibitem [{\citenamefont {{Doneva}}\ \emph {et~al.}(2015)\citenamefont {{Doneva}}, \citenamefont {{Yazadjiev}},\ and\ \citenamefont {{Kokkotas}}}]{doneva2015iq}%
	  \BibitemOpen
	  \bibfield  {author} {\bibinfo {author} {\bibfnamefont {D.~D.}\ \bibnamefont {{Doneva}}}, \bibinfo {author} {\bibfnamefont {S.~S.}\ \bibnamefont {{Yazadjiev}}},\ and\ \bibinfo {author} {\bibfnamefont {K.~D.}\ \bibnamefont {{Kokkotas}}},\ }\bibfield  {title} {\bibinfo {title} {{I-Q relations for rapidly rotating neutron stars in f(R) gravity}},\ }\href {https://doi.org/10.1103/PhysRevD.92.064015} {\bibfield  {journal} {\bibinfo  {journal} {\prd}\ }\textbf {\bibinfo {volume} {92}},\ \bibinfo {eid} {064015} (\bibinfo {year} {2015})}\BibitemShut {NoStop}%
	\bibitem [{\citenamefont {{Gupta}}\ \emph {et~al.}(2018)\citenamefont {{Gupta}}, \citenamefont {{Majumder}}, \citenamefont {{Yagi}},\ and\ \citenamefont {{Yunes}}}]{gupta2018iloveq}%
	  \BibitemOpen
	  \bibfield  {author} {\bibinfo {author} {\bibfnamefont {T.}~\bibnamefont {{Gupta}}}, \bibinfo {author} {\bibfnamefont {B.}~\bibnamefont {{Majumder}}}, \bibinfo {author} {\bibfnamefont {K.}~\bibnamefont {{Yagi}}},\ and\ \bibinfo {author} {\bibfnamefont {N.}~\bibnamefont {{Yunes}}},\ }\bibfield  {title} {\bibinfo {title} {{I-Love-Q relations for neutron stars in dynamical Chern Simons gravity}},\ }\href {https://doi.org/10.1088/1361-6382/aa9c68} {\bibfield  {journal} {\bibinfo  {journal} {Classical and Quantum Gravity}\ }\textbf {\bibinfo {volume} {35}},\ \bibinfo {eid} {025009} (\bibinfo {year} {2018})}\BibitemShut {NoStop}%
	\bibitem [{\citenamefont {Haskell}\ \emph {et~al.}(2013)\citenamefont {Haskell}, \citenamefont {Ciolfi}, \citenamefont {Pannarale},\ and\ \citenamefont {Rezzolla}}]{Haskell:2013}%
	  \BibitemOpen
	  \bibfield  {author} {\bibinfo {author} {\bibfnamefont {B.}~\bibnamefont {Haskell}}, \bibinfo {author} {\bibfnamefont {R.}~\bibnamefont {Ciolfi}}, \bibinfo {author} {\bibfnamefont {F.}~\bibnamefont {Pannarale}},\ and\ \bibinfo {author} {\bibfnamefont {L.}~\bibnamefont {Rezzolla}},\ }\bibfield  {title} {\bibinfo {title} {On the universality of i-love-q relations in magnetized neutron stars},\ }\href {https://doi.org/10.1093/mnrasl/slt161} {\bibfield  {journal} {\bibinfo  {journal} {Mon. Not. R. Astron. Soc.}\ }\textbf {\bibinfo {volume} {438}},\ \bibinfo {pages} {L71–L75} (\bibinfo {year} {2013})}\BibitemShut {NoStop}%
	\bibitem [{\citenamefont {Martinon}\ \emph {et~al.}(2014)\citenamefont {Martinon}, \citenamefont {Maselli}, \citenamefont {Gualtieri},\ and\ \citenamefont {Ferrari}}]{Martinon:2014}%
	  \BibitemOpen
	  \bibfield  {author} {\bibinfo {author} {\bibfnamefont {G.}~\bibnamefont {Martinon}}, \bibinfo {author} {\bibfnamefont {A.}~\bibnamefont {Maselli}}, \bibinfo {author} {\bibfnamefont {L.}~\bibnamefont {Gualtieri}},\ and\ \bibinfo {author} {\bibfnamefont {V.}~\bibnamefont {Ferrari}},\ }\bibfield  {title} {\bibinfo {title} {Rotating protoneutron stars: Spin evolution, maximum mass, and i-love-q relations},\ }\href {https://doi.org/10.1103/PhysRevD.90.064026} {\bibfield  {journal} {\bibinfo  {journal} {Phys. Rev. D}\ }\textbf {\bibinfo {volume} {90}},\ \bibinfo {pages} {064026} (\bibinfo {year} {2014})}\BibitemShut {NoStop}%
	\bibitem [{\citenamefont {Marques}\ \emph {et~al.}(2017)\citenamefont {Marques}, \citenamefont {Oertel}, \citenamefont {Hempel},\ and\ \citenamefont {Novak}}]{Marques:2017}%
	  \BibitemOpen
	  \bibfield  {author} {\bibinfo {author} {\bibfnamefont {M.}~\bibnamefont {Marques}}, \bibinfo {author} {\bibfnamefont {M.}~\bibnamefont {Oertel}}, \bibinfo {author} {\bibfnamefont {M.}~\bibnamefont {Hempel}},\ and\ \bibinfo {author} {\bibfnamefont {J.}~\bibnamefont {Novak}},\ }\bibfield  {title} {\bibinfo {title} {New temperature dependent hyperonic equation of state: Application to rotating neutron star models and $i\text{\ensuremath{-}}q$ relations},\ }\href {https://doi.org/10.1103/PhysRevC.96.045806} {\bibfield  {journal} {\bibinfo  {journal} {Phys. Rev. C}\ }\textbf {\bibinfo {volume} {96}},\ \bibinfo {pages} {045806} (\bibinfo {year} {2017})}\BibitemShut {NoStop}%
	\bibitem [{\citenamefont {{Chavanis}}\ and\ \citenamefont {{Harko}}(2012)}]{chavanis2012bose}%
	  \BibitemOpen
	  \bibfield  {author} {\bibinfo {author} {\bibfnamefont {P.-H.}\ \bibnamefont {{Chavanis}}}\ and\ \bibinfo {author} {\bibfnamefont {T.}~\bibnamefont {{Harko}}},\ }\bibfield  {title} {\bibinfo {title} {{Bose-Einstein condensate general relativistic stars}},\ }\href {https://doi.org/10.1103/PhysRevD.86.064011} {\bibfield  {journal} {\bibinfo  {journal} {\prd}\ }\textbf {\bibinfo {volume} {86}},\ \bibinfo {eid} {064011} (\bibinfo {year} {2012})}\BibitemShut {NoStop}%
	\bibitem [{\citenamefont {{Yeung}}\ \emph {et~al.}(2021)\citenamefont {{Yeung}}, \citenamefont {{Lin}}, \citenamefont {{Andersson}},\ and\ \citenamefont {{Comer}}}]{Yeung:2021}%
	  \BibitemOpen
	  \bibfield  {author} {\bibinfo {author} {\bibfnamefont {C.-H.}\ \bibnamefont {{Yeung}}}, \bibinfo {author} {\bibfnamefont {L.-M.}\ \bibnamefont {{Lin}}}, \bibinfo {author} {\bibfnamefont {N.}~\bibnamefont {{Andersson}}},\ and\ \bibinfo {author} {\bibfnamefont {G.}~\bibnamefont {{Comer}}},\ }\bibfield  {title} {\bibinfo {title} {{The I-Love-Q Relations for Superfluid Neutron Stars}},\ }\href {https://doi.org/10.3390/universe7040111} {\bibfield  {journal} {\bibinfo  {journal} {Universe}\ }\textbf {\bibinfo {volume} {7}},\ \bibinfo {eid} {111} (\bibinfo {year} {2021})},\ \Eprint {https://arxiv.org/abs/2105.00798} {arXiv:2105.00798 [astro-ph.HE]} \BibitemShut {NoStop}%
	\end{thebibliography}

%

\end{document}